\documentclass{aa}  

\usepackage{graphicx}
\usepackage{txfonts}
\usepackage{lipsum}
\usepackage{subcaption}         % necessary for continued figures, example in section 3
\usepackage{lscape}             % to rotate a single page table, example in appendix.
\usepackage{placeins}           % useful with \FloatBarrier, to keep 
\usepackage[colorlinks=true, citecolor=blue]{hyperref}
\usepackage{amsmath}
\usepackage{tablefootnote}

\begin{document}

\title{Elemental abundance pattern and temperature inversion on the dayside of HAT-P-70b observed with CARMENES and PEPSI}
\titlerunning{Elemental abundance pattern and temperature inversion on the dayside of HAT-P-70b}

% \subtitle{Subtitle}

\author{B.~Guo \inst{1,2} \and
        F.~Yan \inst{1} \and 
        Th.~Henning \inst{3} \and
        L.~Nortmann \inst{2} \and
        M.~Stangret \inst{4} \and
        D.~Cont \inst{5,6} \and
        E.~Pall\'e \inst{7,8} \and
        D.~Shulyak \inst{9} \and
        K.\,G.~Strassmeier \inst{10,11} \and
        I.~Ilyin \inst{10} \and
        F.~Lesjak \inst{2,10} \and
        A.~Reiners \inst{2} \and
        S.~Liu \inst{1} \and
        K.~Molaverdikhani \inst{5,6} \and
        G.~Scandariato \inst{12} \and
        E.~Keles \inst{13} \and
        J.\,A.~Caballero \inst{14} \and
        P.\,J.~Amado \inst{9} \and
        A.~Quirrenbach \inst{15} \and
        I.~Ribas \inst{16,17} \and
        S.~G\'ongora \inst{18} \and
        A.\,P.~Hatzes \inst{19} \and
        M.~L\'opez-Puertas \inst{9} \and
        D.~Montes \inst{20} \and
        K.~Poppenhaeger \inst{10,11} \and
        E.~Schlawin \inst{21} \and
        A.~Schweitzer \inst{22} \and
        D.~Sicilia \inst{12}}
\authorrunning{B.~Guo et al.}

\institute{
        %1
        Department of Astronomy, University of Science and Technology of China, Hefei 230026, China\\
        \email{yanfei@ustc.edu.cn} 
        \and 
        %2
        Institut f\"ur Astrophysik und Geophysik, Georg-August-Universit\"at, Friedrich-Hund-Platz 1, 37077 G\"ottingen, Germany 
        \and 
        %3
        Max-Planck-Institut f{\"u}r Astronomie, K{\"o}nigstuhl 17, 69117 Heidelberg, Germany 
        \and
        %4
        INAF – Osservatorio Astronomico di Padova, Vicolo dell'Osservatorio 5, 35122, Padova, Italy 
        \and
        %5
        Universitäts-Sternwarte, Ludwig-Maximilians-Universität München, Scheinerstrasse 1, 81679 München, Germany 
        \and
        %6
        Exzellenzcluster Origins, Boltzmannstrasse 2, 85748 Garching bei München, Germany 
        \and
        %7
        Instituto de Astrof\'isica de Canarias, 38205 La Laguna, Tenerife, Spain
        \and
        %8
        Departamento de Astrof\'isica, Universidad de La Laguna, 38206 La Laguna, Tenerife, Spain
        \and
        %9
        Instituto de Astrofísica de Andalucía (IAA-CSIC), Glorieta de la Astronomía s/n, 18008 Granada, Spain 
        \and
        %10
        Leibniz-Institut f\"ur Astrophysik Potsdam, An der Sternwarte 16, 14482 Potsdam, Germany
        \and
        %11
        Universität Potsdam, Institut für Physik und Astronomie, Karl-Liebknecht-Str. 24/25, 14476 Potsdam-Golm, Germany
        \and
        %12
        INAF – Osservatorio Astrofisico di Catania, Via S. Sofia 78, 95123 Catania, Italy
        \and
        %13
        Freie Universität Berlin, Institute of Geological Sciences, Malteserstrasse 74-100, 12249 Berlin, Germany
        \and
        %14
        Centro de Astrobiolog\'ia (CSIC-INTA), Camino Bajo del Castillo s/n, Campus ESAC, 28692 Villanueva de la Ca\~nada, Madrid, Spain
        \and
        %15
        Landessternwarte, Zentrum f\"ur Astronomie der Universit\"at Heidelberg, K\"onigstuhl 12, 69117 Heidelberg, Germany
        \and
        %16
        Institut de Ci\`encies de l'Espai (CSIC), c/ de Can Magrans s/n, Campus UAB, 08193 Bellaterra, Barcelona, Spain
        \and
        %17
        Institut d'Estudis Espacials de Catalunya, 08860 Castelldefels, Barcelona, Spain
        \and
        %18
        Departamento de F\'isica de la Tierra y Astrof\'isica \& IPARCOS Instituto de F\'isica de Part\'iculas y del Cosmos, Facultad de Ciencias F\'isicas, Universidad Complutense de Madrid, Plaza de Ciencias 1, 28400 Madrid, Spain
        \and
        %19
        Th\"uringer Landessternwarte Tautenburg, Sternwarte 5, 07778 Tautenburg, Germany
        \and
        %20
        Centro Astron\'omico Hispano en Andaluc\'ia, Observatorio Astron\'omico de Calar Alto, Sierra de los Filabres, 04550 G\'ergal, Almer\'ia, Spain \and
        %21
        Steward Observatory, University of Arizona, 933 N. Cherry Ave., Tucson, AZ 85721, USA \and
        %22
        Hamburger Sternwarte, Gojenbergsweg 112, 21029 Hamburg, Germany}

\date{Received 30 October 2025 / Accepted 14 December 2025}

\abstract{Ground-based high-resolution spectroscopic observations have identified various chemical species in the atmosphere of numerous ultra-hot Jupiters (UHJs), including neutral and ionized metals. These detections have offered valuable insights into planet formation mechanisms via abundance measurements of refractory elements. We observed the dayside thermal emission spectrum of UHJ \mbox{HAT-P-70b} using the high-resolution spectrographs CARMENES and PEPSI. Through our cross-correlation analysis, we detected emission signals for \ion{Al}{i}, AlH, \ion{Ca}{ii}, \ion{Cr}{i}, \ion{Fe}{i}, \ion{Fe}{ii}, \ion{Mg}{i}, \ion{Mn}{i}, and \ion{Ti}{i}, marking the first detection of \ion{Al}{i} and AlH in an exoplanetary atmosphere. Tentative signals of \ion{C}{i}, \ion{Ca}{i}, \ion{Na}{i}, NaH, and \ion{Ni}{i} were also identified. Based on those detections, we were able to perform atmospheric retrievals to constrain the thermal profile and elemental abundances of the planet’s dayside hemisphere. The retrieved temperature-pressure profile reveals a strong temperature inversion layer. The chemical free retrieval yielded a metallicity of $\mathrm{[Fe/H]} = 0.38^{+0.74}_{-1.11}$, while the chemical equilibrium retrieval resulted in $\mathrm{[Fe/H]} = 0.23^{+1.08}_{-0.98}$, with both values consistent with the solar metallicity. We also tentatively found an enriched abundance of Ni, which could result from the accretion of Ni-rich planetesimals during the planet’s formation. On the other hand, elements with condensation temperatures above 1400\,K (e.g., Ca, Ti, and V) appear to be slightly depleted, possibly due to cold-trapping on the planet’s nightside. However, Al, with the highest condensation temperature at 1653\,K, displays a solar-like abundance, which might reflect the formation-related enrichment of Al. Our retrieval indicates extremely high volume mixing ratios of metal ions (\ion{Fe}{ii} and \ion{Ca}{ii}), which are significantly inconsistent with predictions from chemical equilibrium models. This disequilibrium suggests that the atmosphere is likely undergoing significant hydrodynamic escaping, which enhances the atmospheric density at high altitudes where the ionic lines are formed.}

\keywords{planets and satellites: atmosphere -- techniques: spectroscopic -- planets and satellites: individuals: HAT-P-70b}

% \titlerunning
\maketitle
\nolinenumbers 
% \longauthor
%%%%%%%%%%%%%%%%%%%%%%%%%%%%%%%%%%%%%%%%%%%%%%%%%%%%%%%%%%%%%%

\section{Introduction}
Ultra-hot Jupiters (UHJs) represent a class of short-period gas giants that suffer extreme irradiation from the host star, typically resulting in high equilibrium temperatures ($T_\mathrm{eq} > 2000\,\mathrm{K}$). In these extreme environments, the dayside hemisphere becomes dominated by atomic and ionic species as most molecules undergo thermal dissociation (e.g., \citealt{inversion_theory2_molecule_dissicoate, molecule_dissociate2, molecule_dissociate3}). The combination of the expanded radius and strong spectral features makes UHJs optimal laboratories for detailed atmospheric characterization. Furthermore, radiative absorption by specific chemical constituents, such as iron, hydrogen, TiO, and VO (e.g., \citealt{inversion_theory1, inversion_theory2_molecule_dissicoate, inversion_theory3, inversion_theory4_kelt9b_nlte}), creates temperature inversion layers where atmospheric temperature increases with altitude. This atmospheric phenomenon has been observationally confirmed in several UHJs, such as KELT-9b, KELT-20b, WASP-178b, and TOI-2109b (e.g., \citealt{kelt9b_intro, kelt-20_yan, wasp-178_cont, toi-2109_cont}).

Recent advances in high-resolution spectroscopy have enabled unprecedented atmospheric characterization of UHJs, with both transmission and thermal emission spectra yielding critical insights. Ground-based high-resolution optical spectrographs, such as  High Dispersion Spectrograph (HDS),  Calar Alto high-Resolution search for M dwarfs with Exoearths with Near-infrared and optical Échelle Spectrographs (CARMENES),  High Accuracy Radial velocity Planet Searcher for the Northern hemisphere (HARPS-N), and  Echelle SpectroPolarimetric Device for the Observation of Stars (ESPaDOnS), have resolved prominent \ion{Fe}{i} signals in UHJs such as WASP-33b (e.g., \citealt{wasp-33_fe1, wasp-33_cont_tio, wasp-33_fe2}), alongside comprehensive detections of neutral and ionized metals (e.g., \ion{V}{i}, \ion{Ti}{ii}, \ion{Ca}{ii}; \citealt{wasp-33_yan_2019, cont_wasp33b_2022, wasp-33_yang_ca}). These findings are complemented by near-infrared (NIR) observations that have simultaneously revealed molecular constituents (e.g., CO, $\mathrm{H_2O}$, OH; \citealt{wasp-33_oh1, wasp-33_yan_co, wasp-33_choi}), providing multiwavelength constraints on atmospheric chemistry. Notably, NIR observations have also provided evidence of nightside emission \citep{yang_wasp33b_nightside, co_wasp33b_nightside}. Additionally, \citet{wasp121_night} also presented the detection of $\mathrm{CH_4}$ in the nightside of WASP-121b with low-resolution JWST (James Webb Space Telescope) spectroscopic observations. 

Bayesian spectral retrieval techniques now enable precise constraints on the atmospheric abundances of detected species. Notable findings include titanium depletion in WASP-76b observed via MAROON-X transmission spectra \citep{wasp76b_cold_trap} attributed to nightside cold-trapping, a phenomenon that has also been identified in WASP-121b \citep{wasp121b_cold_trap}. In contrast, detections of \ion{Ti}{i} and TiO in hotter UHJs such as WASP-33b, WASP-189b, and MASCARA-1b \citep{cont_wasp33b_2022, wasp189b_tio, mascara1b_pepsi, guo_mascara1b} suggest that this condensation mechanism is suppressed at extremely high temperatures. 

In addition, studies of carbon and oxygen abundances primarily focus on constraining the C/O ratio as a tracer of planet formation pathways (e.g., \citealt{C/O_formation_tracker1, C/O_formation_tracker2, C/O_formation_tracker3, C/O_formation_tracker4, C/O_formation_tracker5, C/O_formation_tracker6}). For instance, a near-solar or elevated C/O ratio in WASP-76b \citep{wasp-76_c/o} implies formation outer the CO snow line via gas accretion or between the soot line and the $\mathrm{H_2O}$ snow line. Conversely, the lower C/O ratio and solar to super-solar metallicity of KELT-20b \citep{kelt-20_c/o} are consistent with planet formation beyond the CO snow line, followed by Type~II migration \citep{type_ii_migration}. Nevertheless, current elemental abundance measurements remain ambiguous, unable to definitively determine whether the lower C/O results from oxygen enrichment or carbon depletion. Intrinsic degeneracies persist in linking these ratios to specific protoplanetary disk regions, reflecting the complex interplay of gas and solid-phase accretion processes.

To break the inherent degeneracies in interpreting formation pathways using the C/O ratio alone, complementary measurements of other elemental tracers are required. The abundance of refractory elements (e.g., Si, Fe) in the atmosphere provides unique constraints on planetary formation histories \citep{refractory_to_volatile}. Although atmospheric condensation below 2000\,K increases the probability to obscures refractory elemental signatures in cooler planets, the extreme thermal environments of UHJs preserve these diagnostic tracers. \citet{refractory_formation} demonstrated that combined refractory and volatile abundance analyses can pinpoint formation locations. For WASP-121b, using the high-resolution spectrograph IGRINS, \citet{wasp121b_igrins} suggested that the most likely formation was between the soot line and $\mathrm{H_2O}$ snow line, although formation scenarios between the $\mathrm{H_2O}$ and CO snow lines or beyond the CO snow line remain plausible. Additionally, high-resolution observations with CRIRES$^+$ and ESPRESSO have revealed a volatile-to-refractory enrichment in the atmosphere of WASP-121b, supporting formation in the outer disk near the CO snow line, followed by inward migration \citep{wasp121b_crires}.

In this study, we report the detection of multiple chemical species (e.g., \ion{Al}{i}, AlH, \ion{Fe}{i}, and \ion{Ti}{i}) in the dayside hemisphere of the UHJ HAT-P-70b through high-resolution thermal emission spectroscopy. Using Bayesian atmospheric retrieval methods, we constrain key atmospheric properties, including temperature-pressure profiles and metallicity. Identified by \citet{hatp70b_parameters}, the gas giant planet revolves around an A-type host star with a period of 2.74\,days and exhibits a strongly misaligned orbit (obliquity $\lambda=107.9^{+2.0}_{-1.7}\,\mathrm{degrees}$; \citealt{HATP70b_HARPS}). Previous transmission spectroscopy with HARPS-N \citep{HATP70b_HARPS} revealed atmospheric absorption features from hydrogen Balmer lines, ionized metals (\ion{Ca}{ii}, \ion{Cr}{ii}, \ion{Fe}{ii}), and neutral species (\ion{Cr}{i}, \ion{Fe}{i}, \ion{Mg}{i}, \ion{Na}{i}, \ion{V}{i}), with tentative detections of \ion{Ca}{i} and \ion{Ti}{ii}. \citet{hatp70b_ghost} also identified the \ion{Ca}{ii} triplet lines in the transmission spectra with GHOST. Our work presents the first thermal emission characterization of this exoplanet's dayside hemisphere. 

This paper is structured as follows. The observational methodology and data reduction are presented in Sect.~\ref{observation_and_data_reduction}. The detection techniques for atmospheric composition, along with their corresponding results are detailed in Sect.~\ref{atmospheric_composition_detection}. We describe the atmospheric retrieval framework and discuss the retrieval results in Sect.~\ref{atmospheric_retrieval}. The conclusions of our work are presented in Sect.~\ref{conclusions}.
%%%%%%%%%%%%%%%%%%%%%%%%%%%%%%%%%%%%%%%%%%%%%%%%%%%%%%%%%%%%%%

\section{Observations and data reduction}
\label{observation_and_data_reduction}

We observed the thermal emission spectra of HAT-P-70b's dayside hemisphere on November 23, 2020 (program ID: H20-3.5-015) and December 23, 2023 (as part of the CARMENES Legacy program), with the CARMENES high-resolution spectrograph \citep{carmenes_spectrograph} installed at the 3.5m telescope of the Observatorio Astron\'omico de Calar Alto in Almer\'ia, Spain. Both visible (VIS, 5200--9600\,$\text{\AA}$, $R$\,\textasciitilde\,94\,600) and NIR (9600--17\,100\,$\text{\AA}$, $R$\,\textasciitilde\,80\,400) wavelength channels were used to observe the target simultaneously. In addition, thermal emission spectra of the UHJ were also observed during the two nights of November 20, 2021 and December 9, 2021, with the Potsdam Echelle Polarimetric and Spectroscopic Instrument (PEPSI, \citealt{pepsi_spectrograph2015, pepsi_spectrograph2018}) mounted on the Large Binocular Telescope (LBT) at the Mount Graham International Observatory in Arizona, USA. These two nights covered similar planetary orbital to those observed with CARMENES, corresponding to before and after eclipse, respectively. During each observation night, we used the LBT in binocular mode, utilizing the cross-disperser (CD) of the PEPSI spectrograph with CD3 in the blue arm (4760--5430\,$\text{\AA}$, $R$\,\textasciitilde\,50\,000) and CD6 in the red arm (7360--8900\,$\text{\AA}$, $R$\,\textasciitilde\,50\,000). These observations were carried out as part of the PEPSI exoplanet transit survey (PETS), and this work is the sixth publication of the PETS series \citep{PETS_I, PETS_II, mascara1b_pepsi, PETS_IV, PETS_V}. The detailed observation logs are presented in Table~\ref{observe_log}.

The raw CARMENES data were reduced using the \texttt{CARACAL} pipeline \citep{carmenes_pipeline_1, carmenes_pipeline_2}, which performs standard calibration including dark subtraction, flat-field correction, wavelength calibration, and spectrum extraction. The pipeline produced two-dimensional (2D) spectra containing 61 spectral orders in the VIS channel and 28 orders in the NIR channel. Additionally, the noise estimates for each spectral data point were also directly obtained from the pipeline. The reduction of PEPSI data was performed on the Spectroscopic Data Systems \citep{sds1,sds2}, a generic software platform capable of fully processing PEPSI data from raw input to final output automatically without human intervention. The reduction process primarily included bias subtraction, flat-field correction, scattered light removal, wavelength calibration, and spectrum normalization. Ultimately, for the PEPSI data, we obtained order-merged spectra.

Initially, the original spectra were cleaned by masking the wavelength points with low signal-to-noise ratio (S/N). For the CARMENES data, a uniform S/N threshold of 30 was applied across all observations. For the PEPSI observations, the thresholds varied between nights and spectral arms: on the first night, we applied S/N cuts of 200 for the blue arm and 175 for the red arm, while on the second night a uniform threshold of 130 was used for both arms. In addition, for the red part of the PEPSI spectra, we also excluded wavelength regions affected by telluric $\mathrm{O_2}$ absorption lines (7590–7700\,$\text{\AA}$, 8100–8370\,$\text{\AA}$). The masked data points accounted for approximately 30\% of the total number of pixels. We then normalized the spectra using a seventh-order polynomial fit and applied a 5$\sigma$ clipping to remove potential outliers, such as strong sky emission lines. For the CARMENES spectra, all spectral orders were further combined to produce one-dimensional (1D) spectra.

To eliminate the contributions of telluric and stellar lines, we applied the \texttt{SYSREM} algorithm \citep{sysrem} to correct for systematic effects. This principal component analysis approach requires no prior assumptions, operating directly on the observed spectral matrix and corresponding noise estimates. The algorithm iteratively optimizes temporal and spectral correction vectors to construct the \texttt{SYSREM} model, which captures time- and wavelength-dependent systematics. We performed ten iterations of this process on the normalized spectra, progressively refining systematic effect removal. The final corrected spectra were obtained by dividing the normalized spectra by the \texttt{SYSREM} model, yielding residual spectra ultimately used for cross-correlation analysis. Noise was calculated throughout the procedure using standard error propagation techniques. A detailed mathematical description of the \texttt{SYSREM} algorithm can be found in \citet{SC_2024}.

\begin{table*}
\caption{Observation logs.}
\label{observe_log} 
\centering
{\renewcommand{\arraystretch}{1.2}
\begin{tabular}{lcccccc}
\hline\hline       
& Date & Airmass change & Exposure time\,(s) & $N_\mathrm{spectra}$ & Phase coverage & S/N range\tablefootmark{(a)} \\
\hline
CARMENES \\
Night 1 & 2020-11-23 & 1.18--1.12--1.55 & 400 & 34 & 0.527--0.592 & 74--87 \\
Night 2 & 2023-12-23 & 1.31--1.12--1.23 & 400 & 31 & 0.419--0.479 & 50--71 \\
\hline
PEPSI \\
Night 1 & 2021-11-20 & 1.28--1.08--1.19 & 300 & 45 & 0.530--0.590 & 209--233 \\
Night 2 & 2021-12-09 & 1.60--1.08--1.10 & 300 & 45 & 0.418--0.478 & 134--175 \\

\hline
\end{tabular}}
\tablefoot{\tablefoottext{a}{The S/N per pixel was measured at \textasciitilde\,6245\,$\text{\AA}$ for CARMENES data and \textasciitilde\,5061\,$\text{\AA}$ for PEPSI data.}}
\end{table*}

%%%%%%%%%%%%%%%%%%%%%%%%%%%%%%%%%%%%%%%%%%%%%%%%%%%%%%%%%%%%%%

\section{Atmospheric composition detection}
\label{atmospheric_composition_detection}
\subsection{Detection methods}
\label{detection_methods}
\subsubsection{Model spectra}

Prior to conducting cross-correlation analysis, we generated thermal emission spectral models for each chemical species in the atmosphere of HAT-P-70b. The model spectra were calculated in a manner similar to that described by \citet{yan_2020_wasp189b}, employing a two-point temperature-pressure ($T$-$P$) profile with the higher altitude point ($T_1$, $P_1$) set at $T_1 = 4500\,\mathrm{K}$, $P_1=10^{-4}\,\mathrm{bar}$, and the lower altitude point ($T_2$, $P_2$) set at $T_2 = 2500\,\mathrm{K}$, $P_2=10^{-2}\,\mathrm{bar}$. The mixing ratio of each chemical species was assumed to be constant throughout the entire $T$–$P$ profile and set to the solar abundance value. A full list of the investigated chemical species, as well as the opacity databases used, is provided in Table~\ref{opacity}. All parameters used in model calculation are summarized in Table~\ref{planetary_parameters}. Since the mass of \mbox{HAT-P-70b} has not been precisely determined, with only an upper limit $M_\mathrm{p} < 6.78\,M_\mathrm{Jup}$ reported by \citet{hatp70b_parameters}, and given prior detections of atmospheric signatures via planetary transmission spectra, we adopted a surface gravity of $\log{g}=3.0\,\mathrm{(cgs)}$ during model spectra calculation and following atmospheric retrievals. This choice was motivated by the requirement for a relatively low surface gravity, which would result in a larger atmospheric scale height, thereby producing strong atmospheric signals detectable by transmission spectroscopy.

We computed the planetary thermal emission spectra using the radiative transfer code \texttt{petitRADTRANS} \citep{pRT_2019}. To ensure consistency with the data reduction processing, we normalized the model spectra by dividing them with the black body radiation spectra of the host star, calculated at its effective temperature. The normalized spectra were then convolved with a Gaussian instrumental profile matching the resolution of the observing instrument, using the \texttt{broadGaussFast} code from \texttt{PyAstronomy} \citep{pyastronomy}. The final model spectra for several chemical species we used in the cross-correlation analysis are presented in Figs.~\ref{Kpmap_signal1}, \ref{Kpmap_signal2}, and \ref{Kpmap_tentative}.

\begin{table}
\centering
\caption{Parameters of the HAT-P-70 system.}
\label{planetary_parameters}
{\renewcommand{\arraystretch}{1.2}
\begin{tabular}{lc}
\hline \hline
Parameter (Unit) & Value\tablefootmark{(a)} \\
\hline
\textit{The star} \\
Radius $R_\star$\,($R_\odot$) & $1.858^{+0.119}_{-0.091}$ \\
Mass $M_\star$\,($M_\odot$) & $1.890^{+0.010}_{-0.013}$ \\
Magnitude $V$\,(mag) & $9.470\,\pm\,0.004$ \\
Effective temperature $T_\mathrm{eff}$\,(K) & $8450^{+540}_{-690}$ \\
Systemic velocity $v_\mathrm{sys}$\,($\mathrm{km\,s^{-1}}$) & $25.26\,\pm\,0.11$ \\
Rotational velocity $v\sin i_\star$\,($\mathrm{km\,s^{-1}}$) & $99.85^{+0.64}_{-0.61}$ \\
Metallicity [Fe/H] & $-0.059^{+0.075}_{-0.088}$ \\
\hline
\textit{The planet}\\
Radius $R_\mathrm{p}$\,($R_\mathrm{Jup}$) & $1.87^{+0.15}_{-0.10}$ \\
Mass $M_\mathrm{p}$\,($M_\mathrm{Jup}$) & $< 6.78\,(3\sigma)$ \\
Surface gravity $\log{g}$\,($\mathrm{log\,cgs}$) & $< 3.73\,(3\sigma)$ \\
Equilibrium temperature $T_\mathrm{eq}$\,(K)\tablefootmark{(b)} & $2562^{+43}_{-52}$ \\
Inclination $i_\mathrm{p}$\,(deg) & $96.50^{+1.42}_{-0.91}$ \\
Orbital period $P$\,(d) & $2.74432452^{+0.00000079}_{-0.00000068}$ \\
Transit epoch $T_0$\,(BJD--TDB) & $2458439.57519^{+0.00045}_{-0.00037}$ \\
Transit duration $T_{14}$\,(h) & $3.480^{+0.067}_{-0.048}$ \\
RV semi-amplitude $K_\mathrm{p}$\,($\mathrm{km\,s^{-1}}$) & $186.82^{+0.33}_{-0.43}$ \\
\hline
\end{tabular}}

\tablefoot{\tablefoottext{a}{All parameters except $K_\mathrm{p}$, which was calculated by Eq.~(\ref{Kp_calculate}), were adopted from \citet{hatp70b_parameters}.}\tablefoottext{b}{$T_\mathrm{eq}$ calculated assuming 0 Bond albedo and full heat redistribution.}}
\end{table}

\subsubsection{Accounting for \texttt{SYSREM} effects}
\label{account_sysrem}
During the data reduction process, we employed the \texttt{SYSREM} algorithm to eliminate the telluric and stellar lines (Sect.~\ref{observation_and_data_reduction}). This procedure can potentially affect the strength and shape of the planetary spectral lines in the data. Accounting for this effect by applying the identical \texttt{SYSREM} processing to the spectral model is expected to improve the detection capability of the cross-correlation method. Our results demonstrate (compared to the conventional approach that does not account for line distortions by \texttt{SYSREM}) that this method enhances the detection significance of the \ion{Fe}{i} signal by approximately 1$\sigma$. The improvement was particularly pronounced for the higher resolution CARMENES observations. 

Given the phase-dependent distortion of line profiles, we initially Doppler-shifted the model spectra into the planetary rest frame with the theoretical radial velocity (RV) at each orbital phase, yielding a phase-resolved 2D model spectral matrix in the planetary rest frame:
\begin{align}
    v_\mathrm{p} &= v_\mathrm{sys} - v_\mathrm{bary} + K_\mathrm{p}\sin{2\pi\phi}, \label{vp_calculate} \\
    K_\mathrm{p} &= \left(\frac{2\pi GM_\star}{P}\right)^{\frac{1}{3}}\sin i_\mathrm{p}, \label{Kp_calculate}
\end{align}
where $v_\mathrm{sys}$ is the systematic velocity, $v_\mathrm{bary}$ is the barycentric velocity of the observer on the Earth, $\phi$ is the planetary orbital phase ($\phi = 0$ corresponds to the middle transit time), $G$ is the gravitational constant, $P$ is the orbital period, $M_\star$ is the stellar mass, and $i_\mathrm{p}$ is the orbital inclination, with all adopted parameter values summarized in Table~\ref{planetary_parameters}. We utilized the accelerated \texttt{SYSREM} framework described by \citet{fast_sysrem} to optimize computational efficiency. Following this approach, during the application of \texttt{SYSREM} to the observed spectra, a corresponding filter matrix was simultaneously generated. The filter was then applied to the model spectral matrix, resulting in a SYSREM-processed model spectral matrix. A model spectral grid was then generated by Doppler-shifting the model across RV from -250\,$\mathrm{km\,s^{-1}}$ to 250\,$\mathrm{km\,s^{-1}}$ in 1\,$\mathrm{km\,s^{-1}}$ increments, which were ready for the final cross-correlation analysis.

\subsubsection{Cross-correlation}
The cross-correlation function (CCF) was computed as the sum of the products of the model and residual spectra, weighted at each wavelength point by the inverse square of the observational noise (1/$\sigma^2$). For a given orbital phase, $\phi$, and velocity shift, $v$, the CCF was calculated as
\begin{equation}
\mathrm{CCF}(v, \phi) = \sum_{N_\lambda}\frac{r(\phi, \lambda)\,m(v, \phi, \lambda)}{\sigma(\phi, \lambda)^2}, 
\label{ccf_calculate}
\end{equation}
where $r(\phi, \lambda)$ and $\sigma(\phi, \lambda)$ represent the residual spectra and their corresponding noise, respectively, $m(v, \phi, \lambda)$ denotes the model spectra, and the summation was performed over all $N_\lambda$ wavelength points.

\begin{figure}
\centering
\includegraphics[width=\hsize]{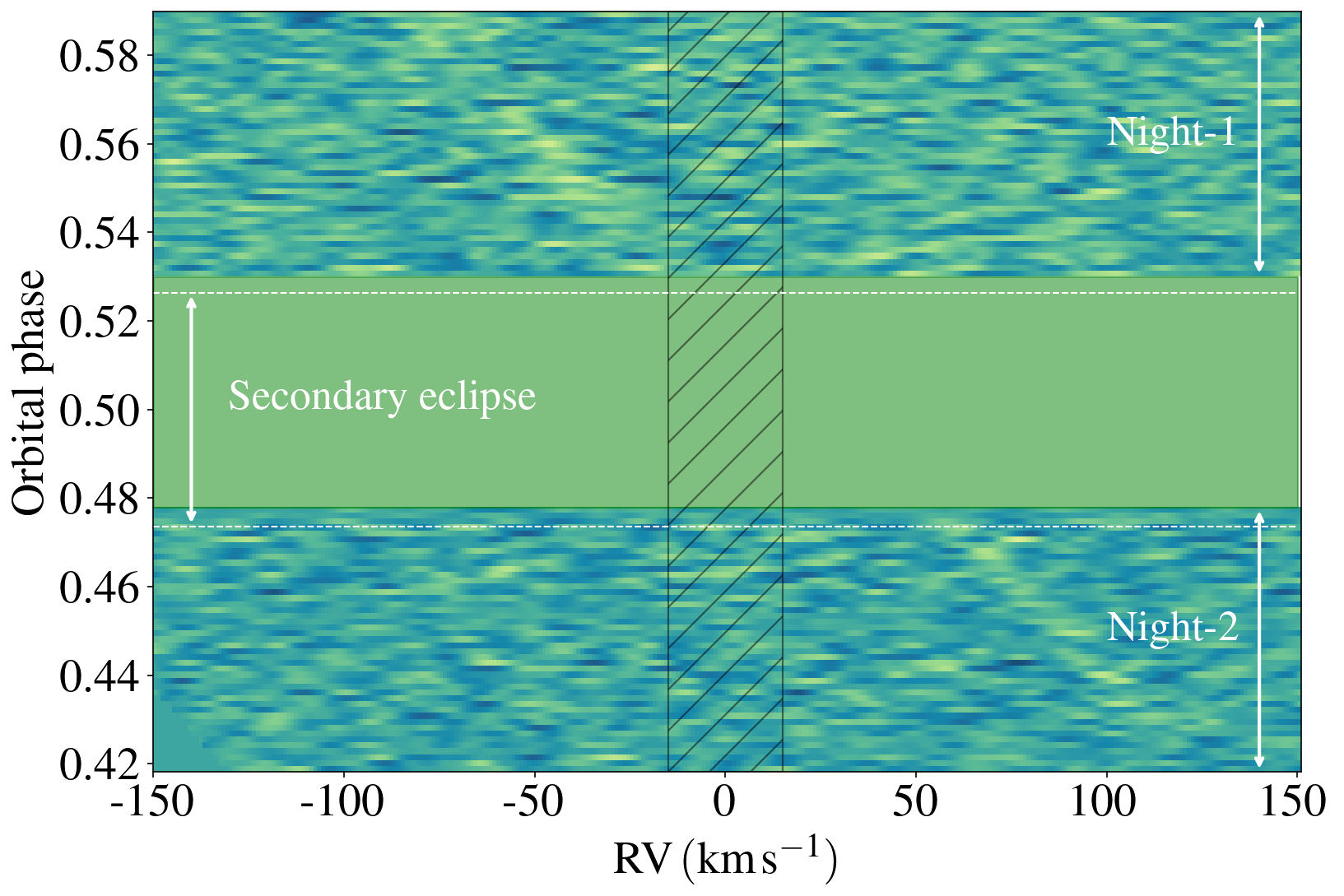}
\caption{CCF for \ion{Fe}{i} in the observer's rest frame, obtained from PEPSI observations. Horizontal dashed lines indicate the temporal boundaries of the secondary eclipse. The hatched region highlights the moonlight-contaminated zone, which was systematically masked during the construction of the $K_\mathrm{p}$--$\Delta v$ map.}
\label{ccf_map_Fe}
\end{figure}

The PEPSI spectra were contaminated by scattered moonlight during the first night, when the Moon-target angular separation was only about 14\,deg and the observation occurred close to the full moon. To mitigate this effect, we excluded CCF within the RV range of $-15<\mathrm{RV}<15\,\mathrm{km\,s^{-1}}$ in the observer's rest frame, as the moonlight contamination contains reflected sunlight whose spectral lines are nearly consistent with the observer’s frame. Figure~\ref{ccf_map_Fe} shows the resulting CCF map for \ion{Fe}{i} obtained from PEPSI data. Although \texttt{SYSREM} processing suppressed much of this contamination, making it not visually prominent in the final CCF map, we found that excluding this RV range still marginally increased the detection significance compared to the unmasked case. Therefore, this masking procedure was retained. In contrast, the CARMENES observations remained unaffected by lunar contamination and therefore required no additional processing.

The CCFs were initially calculated in the observer’s rest frame. To search for the planetary atmosphere signal, we transformed the CCFs into the planetary rest frame using Eq.~(\ref{vp_calculate}), by applying a range of trial values for $K_\mathrm{p}$, ranging from 100 to 300\,$\mathrm{km\,s^{-1}}$ in 1\,$\mathrm{km\,s^{-1}}$ increments. For each trial $K_\mathrm{p}$, the CCFs were integrated along the orbital phase axis to yield a 1D array. The horizontal axis of this array corresponds to the RV relative to the planetary rest frame ($\mathrm{RV}=0$), which we denote as $\Delta v$. The parameter $\Delta v$ quantifies the uncertainty of $v_\mathrm{sys}$, while also provides critical insights into planetary rotation and atmospheric circulation patterns. The ensemble of 1D arrays from all $K_\mathrm{p}$ trials was stacked to form a 2D parameter map ($K_\mathrm{p}$--$\Delta v$ map). To evaluate the detection significance, we constructed the S/N map by normalizing a $K_\mathrm{p}$--$\Delta v$ map for each species with the standard deviation of the noise-dominated regions ($50\,\mathrm{km\,s^{-1}} < |\Delta v| < 100\,\mathrm{km\,s^{-1}}$ and $100\,\mathrm{km\,s^{-1}} < K_\mathrm{p} < 300\,\mathrm{km\,s^{-1}}$), excluding the central signal peak. A significant signal near the theoretical $K_\mathrm{p}$ and $\Delta v = 0$ in the S/N map provides conclusive evidence of the presence of the chemical species in the planetary atmosphere.

\subsection{Detection results}

\begin{figure*}
\centering
\includegraphics[width=\textwidth]{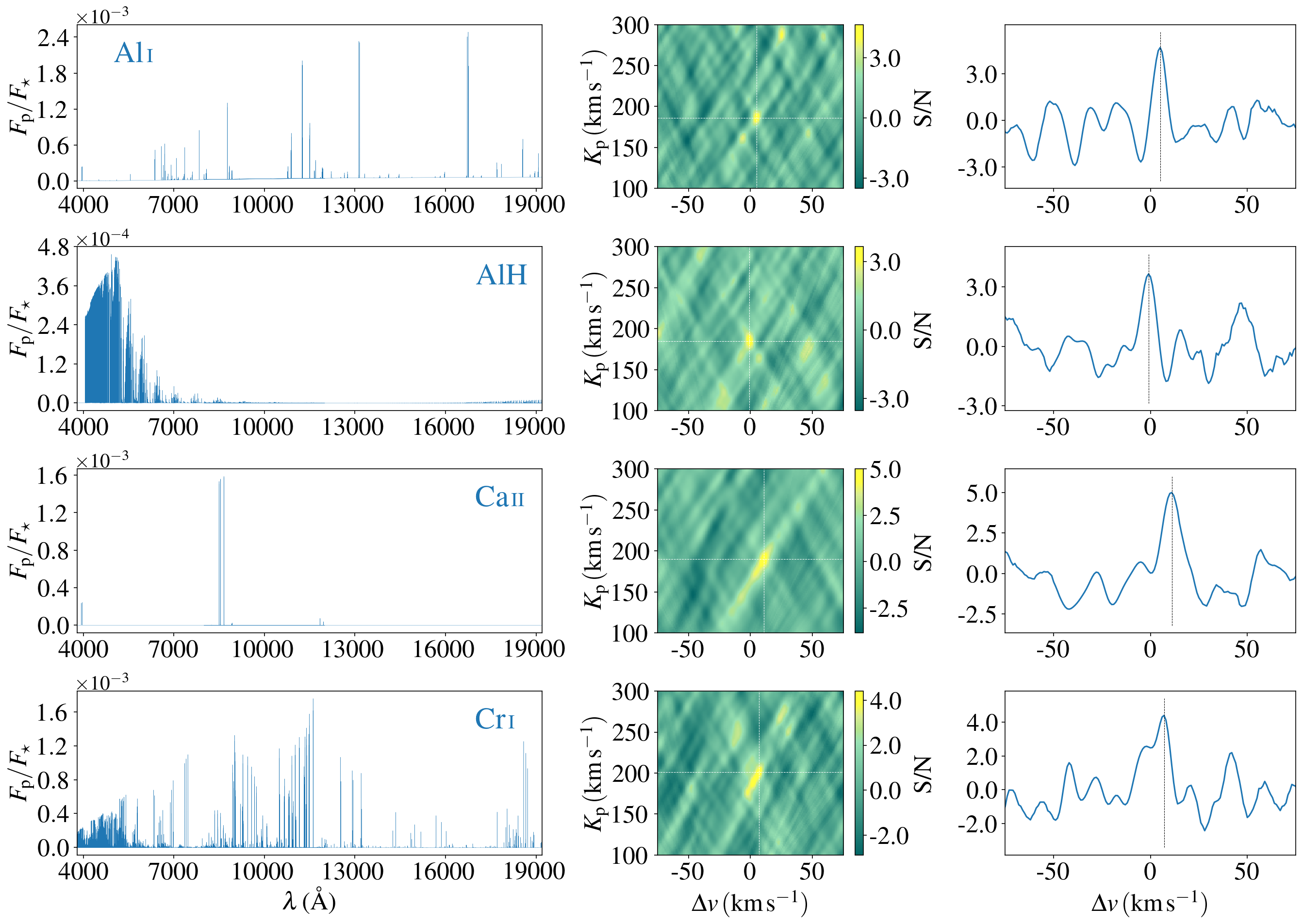}
\caption{Model spectra and S/N maps for \ion{Al}{i}, AlH, \ion{Ca}{ii}, and \ion{Cr}{i}. \textit{Left panels}: Model spectra for each chemical species. \textit{Middle panels}: Corresponding S/N maps, combining data from both CARMENES and PEPSI instruments. The white dotted lines indicate the position of $K_\mathrm{p}$--$\Delta v$ where the S/N reaches its maximum. \textit{Right panels}: CCFs at the $K_\mathrm{p}$ values corresponding to the maximum S/N detections.}
\label{Kpmap_signal1}
\end{figure*}

\begin{figure*}
\centering
\includegraphics[width=\textwidth]{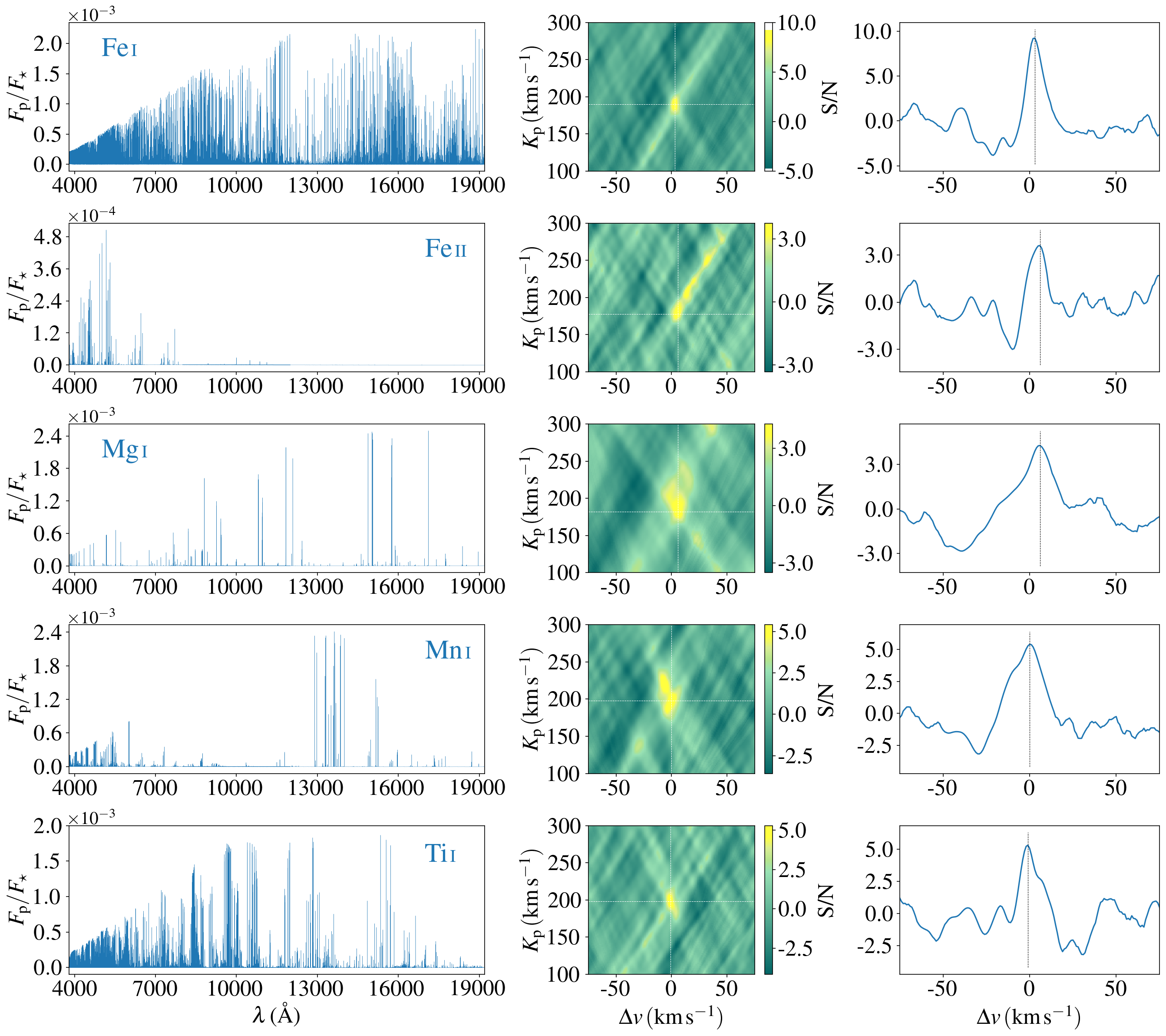}
\caption{Same as Fig.~\ref{Kpmap_signal1}, but for \ion{Fe}{i}, \ion{Fe}{ii}, \ion{Mg}{i}, \ion{Mn}{i}, and \ion{Ti}{i}.}
\label{Kpmap_signal2}
\end{figure*}

\begin{figure*}
\centering
\includegraphics[width=\textwidth]{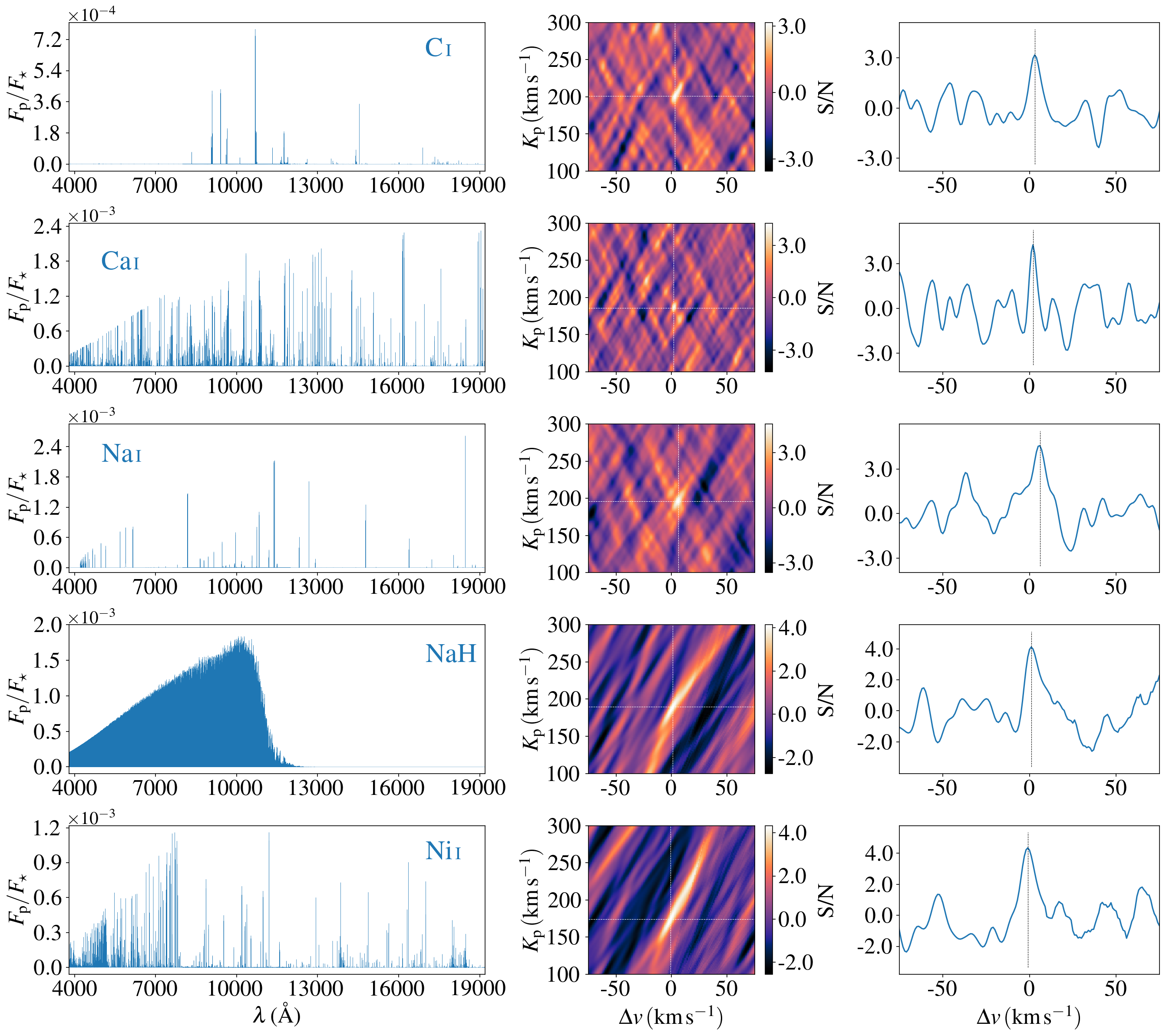}
\caption{Same as Fig.~\ref{Kpmap_signal1} but showing tentative detections. The S/N maps for \ion{Ca}{i} and \ion{Na}{i} are derived exclusively from CARMENES spectra, whereas those for NaH and \ion{Ni}{i} only use  the first night  data.}
\label{Kpmap_tentative}
\end{figure*}

We detected the atmospheric emission signals of \ion{Al}{i}, AlH, \ion{Ca}{ii}, \ion{Cr}{i}, \ion{Fe}{i}, \ion{Fe}{ii}, \ion{Mg}{i}, \ion{Mn}{i}, and \ion{Ti}{i}. The corresponding S/N maps are illustrated in Figs.~\ref{Kpmap_signal1} and \ref{Kpmap_signal2}, which combine all observations from CARMENES and PEPSI. We combined the $K_\mathrm{p}$--$\Delta v$ maps from all observations and subsequently normalized the result to obtain the final S/N map. This procedure effectively corresponds to a weighted summation, with the final map being predominantly dominated by the PEPSI observations with higher S/N. For each species, we selected the \texttt{SYSREM} iteration that yields the highest significance value in the S/N map. Figure~\ref{SNR_iter} illustrates the variation of the \ion{Fe}{i} signal S/N with the number of \texttt{SYSREM} iterations. Other species exhibit a similar behavior, with their S/N values converging after approximately two iterations. A summary of the detection results is provided in Table~\ref{detection_results_table}.

\begin{table}
\centering
\caption{Summary of the cross-correlation results.}
\label{detection_results_table}
{\renewcommand{\arraystretch}{1.2}
\begin{tabular}{lccc}
\hline \hline
Species & S/N & $K_\mathrm{p}\,(\mathrm{km\,s^{-1}})$ & $\Delta v\,(\mathrm{km\,s^{-1}})$ \\
\hline
\textit{Detected signals} \\
\ion{Al}{i} & 4.7 & $186.0_{-6.0}^{+5.7}$ & $4.9_{-2.3}^{+2.1}$ \\
AlH & 3.7 & $185.2\pm8.3$ & $-1.0_{-2.8}^{+2.7}$ \\
\ion{Ca}{ii} & 5.0 & $189.9_{-20.0}^{+8.0}$ & $10.7_{-7.5}^{+3.2}$ \\
\ion{Cr}{i} & 4.4 & $200.7_{-28.4}^{+7.5}$ & $6.7_{-10.7}^{+2.6}$ \\
\ion{Fe}{i} & 9.3 & $189.8_{-6.5}^{+6.1}$ & $2.6_{-2.0}^{+2.1}$ \\
\ion{Fe}{ii} & 3.8 & $178.1_{-8.6}^{+16.2}$ & $5.6_{-4.4}^{+4.3}$ \\
\ion{Mg}{i} & 4.3 & $182.3_{-10.5}^{+31.2}$ & $5.6_{-10.4}^{+6.4}$ \\
\ion{Mn}{i} & 5.4 & $198.2_{-17.5}^{+35.1}$ & $0.4_{-10.6}^{+4.3}$ \\
\ion{Ti}{i} & 5.3 & $198.3_{-9.5}^{+9.2}$ & $-1.3_{-2.7}^{+3.4}$ \\
\textit{Tentative signals} \\
\ion{C}{i} & 3.2 & $200.7_{-8.4}^{+17.4}$ & $3.0_{-2.9}^{+6.4}$ \\
\ion{Ca}{i} & 4.3 & $186.4_{-4.1}^{+4.7}$ & $2.0_{-1.3}^{+1.6}$ \\
\ion{Na}{i} & 4.6 & $196.1_{-9.4}^{+10.0}$ & $5.6_{-3.9}^{+4.0}$ \\
NaH & 4.1 & $187.9_{-22.3}^{+45.2}$ & $0.3_{-8.1}^{+20.7}$ \\
\ion{Ni}{i} & 4.4 & $172.5_{-22.9}^{+80.9}$ & $-1.5_{-8.1}^{+26.5}$ \\
\hline
\end{tabular}}
\end{table}

\begin{figure}
\centering
\includegraphics[width=\hsize]{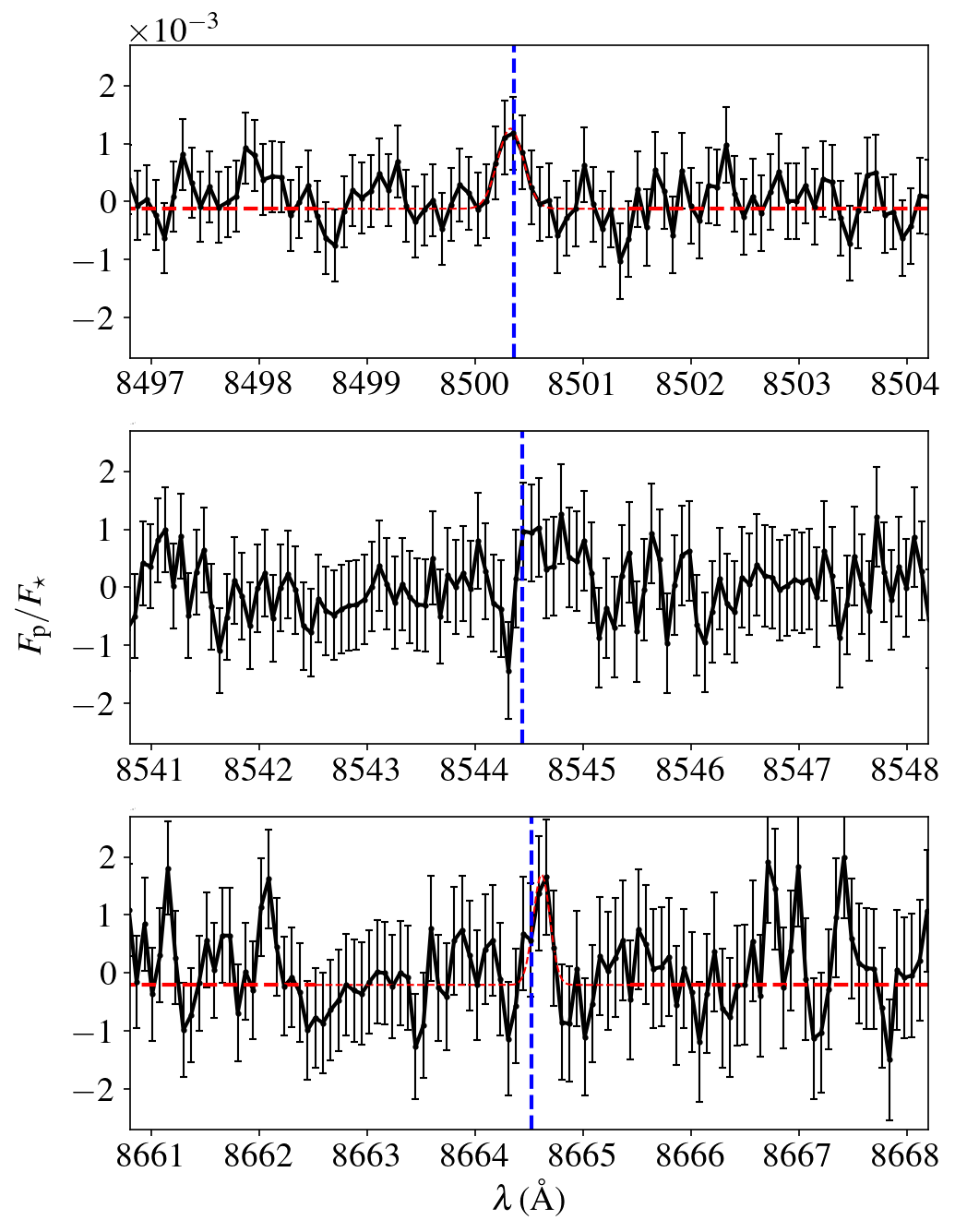}
\caption{\ion{Ca}{ii} triplet emission lines observed with PEPSI. They were combined over two nights and shifted to the planetary rest frame using the best-fit velocity from the \ion{Ca}{ii} CCF signal. The dashed blue lines indicate the expected positions of the \ion{Ca}{ii} triplet lines. Among them, only the lines near 8500.36\,$\mathrm{\AA}$ and 8664.52\,$\mathrm{\AA}$ (vacuum wavelengths) were detected, with their Gaussian fits in red.}
\label{Ca_signal_line}
\end{figure}

This work presents the first thermal emission observation of HAT-P-70b, and reports the first detection of \ion{Al}{i}, AlH, \ion{Mn}{i}, and \ion{Ti}{i} in its atmosphere. The detections of \ion{Ca}{ii}, \ion{Cr}{i}, \ion{Fe}{i}, \ion{Fe}{ii}, and \ion{Mg}{i} are consistent with previous transmission spectroscopy results from HARPS-N \citep{HATP70b_HARPS}. To further validate the \ion{Ca}{ii} signal, we inspected its individual triplet lines in the residual spectra (Fig.~\ref{Ca_signal_line}). We detected the first and third lines of the \ion{Ca}{ii} triplet, which independently confirms the detection of \ion{Ca}{ii} previously reported from GHOST transmission observations \citep{hatp70b_ghost}.

General velocity offsets are observed among the detected signals. The $K_\mathrm{p}$ values for \ion{Al}{i}, AlH, \ion{Ca}{ii}, and \ion{Fe}{i} cluster around 190\,$\mathrm{km\,s^{-1}}$, closely matching the theoretically expected value of $186.82^{+0.33}_{-0.43}\,\mathrm{km\,s^{-1}}$. Conversely, \ion{Cr}{i}, \ion{Mn}{i}, and \ion{Ti}{i} yield $K_\mathrm{p}$ near 200\,$\mathrm{km\,s^{-1}}$, whereas \ion{Fe}{ii} and \ion{Mg}{i} are detected around 180\,$\mathrm{km\,s^{-1}}$. Despite these variations, the large measurement uncertainties of 6--35\,$\mathrm{km\,s^{-1}}$ render all observed $K_\mathrm{p}$ values consistent with the theoretical prediction within 1$\sigma$ range. Therefore, we regarded these detections as robust. Velocity deviations in $K_\mathrm{p}$ could arise from several physical mechanisms, including 3D atmospheric structure, atmospheric dynamics, and planetary rotation (e.g., \citealt{v_offset}). Similarly, $\Delta v$ offsets can be induced by uncertainties in $v_\mathrm{sys}$ and atmospheric dynamics such as day-to-night winds. Nevertheless, given the substantial uncertainties in our detection, the presence of these offsets cannot be definitively confirmed from the current observations.

\begin{figure}
\centering
\includegraphics[width=\hsize]{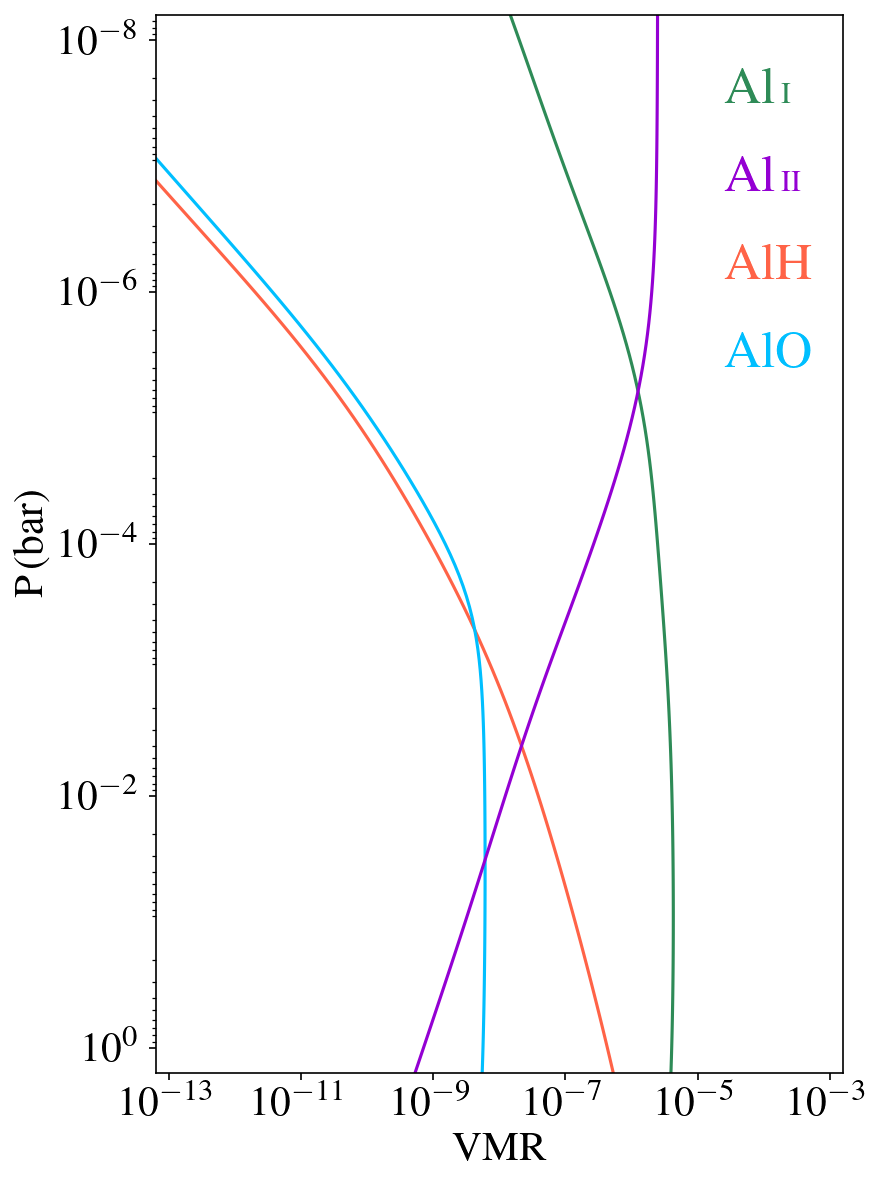}
\caption{Volume mixing ratios of \ion{Al}{i}, \ion{Al}{ii}, AlO, and AlH as a function of pressure, calculated under the assumption of chemical equilibrium. An isothermal temperature profile of 2500\,K was adopted.}
\label{Al_eq}
\end{figure}

This work presents the first detection of \ion{Al}{i} and AlH in an exoplanet's atmosphere. Previous searches for aluminum had primarily focused on AlO, a species considered a potential condensate forming clouds or hazes in hot exoplanet atmospheres (e.g., \citealt{alo_condense1, alo_condense2, alo_condense3}). \citet{alo_inversion} also proposed that the presence of AlO in the planetary atmosphere may enhance visible opacity, leading to the absorption of stellar irradiation and the formation of a temperature inversion layer. For instance, \citet{alo_wasp33} reported the first tentative detection (3.3$\sigma$) of AlO in the UHJ WASP-33b with low-resolution transmission spectra from OSIRIS at the Gran Telescopio Canarias. Subsequently, \citet{alo_wasp43}, through a re-analysis of {\em Hubble} WFC3 transmission spectra combined with {\em Spitzer} data, confirmed the presence of AlO (>$5\sigma$) in WASP-43b, interpreting it as a potential evidence of the disequilibrium processes, such as vertical or horizontal mixing. Additionally, \citet{alo_wasp17} also reported a potential existence of AlO in WASP-17b using the {\em Hubble} STIS and WFC3 transmission spectra. However, at the high temperatures characteristic of UHJs, chemical equilibrium models assuming solar abundances predict that \ion{Al}{i} and AlH should be significantly more abundant than AlO (Fig.~\ref{Al_eq}). Therefore, future atmospheric characterization efforts targeting aluminum in UHJs should arguably prioritize searches for \ion{Al}{i} and AlH, given their expected prominence in these extreme environments.

Due to the high condensation temperature of aluminum ($T_\mathrm{cond,50\%} = 1653\,\mathrm{K}$, \citealt{cond_temp}) and titanium ($T_\mathrm{cond,50\%} = 1582\,\mathrm{K}$), species containing these elements are expected to predominantly condense and become cold-trapped in the cooler nightside of hot Jupiters or UHJs. For example, the non-detection of Ti in the atmospheres of WASP-76b and \mbox{WASP-121b}, UHJs with lower equilibrium temperatures (WASP-76b: $T_\mathrm{eq}=2231^{+37}_{-36}\,\mathrm{K}$, \citealt{wasp76_teq}; \mbox{WASP-121b}: $T_\mathrm{eq}=2358\pm52\,\mathrm{K}$, \citealt{wasp121_teq}), has been attributed to this type of nightside cold trap. Conversely, several studies have reported detections of \ion{Ti}{i} and TiO in hotter UHJs \citep{cont_wasp33b_2022, wasp189b_tio, mascara1b_pepsi, guo_mascara1b}, such as \mbox{WASP-33b}, \mbox{WASP-189b}, and \mbox{MASCARA-1b} (\mbox{WASP-33b}: $T_\mathrm{eq}=2781.7\pm41.1\,\mathrm{K}$, \citealt{wasp33_teq}; \mbox{WASP-189b}: $T_\mathrm{eq}=3353^{+27}_{-34}\,\mathrm{K}$, \citealt{wasp189_teq}; \mbox{MASCARA-1b}: $T_\mathrm{eq}=2594.3^{+1.6}_{-1.5}\,\mathrm{K}$, \citealt{mascara1_teq}). In addition, \citet{ti_hd149026} also successfully detected \ion{Ti}{i} in \mbox{HD 149026b}, a hot Saturn with $T_\mathrm{eq}=1626^{+69}_{-37}\,\mathrm{K}$ \citep{149026_teq}, which was interpreted as resulting from both efficient energy redistribution across the planet and a low Bond albedo ($A\sim0.1$). Thus, the detection of \ion{Al}{i}, AlH, and \ion{Ti}{i} in the dayside hemisphere of HAT-P-70b may be attributed to a combination of factors: its high equilibrium temperature and potentially efficient energy and materials transport between the planetary day and night sides hemisphere.

\begin{figure*}
\centering
\includegraphics[width=\textwidth]{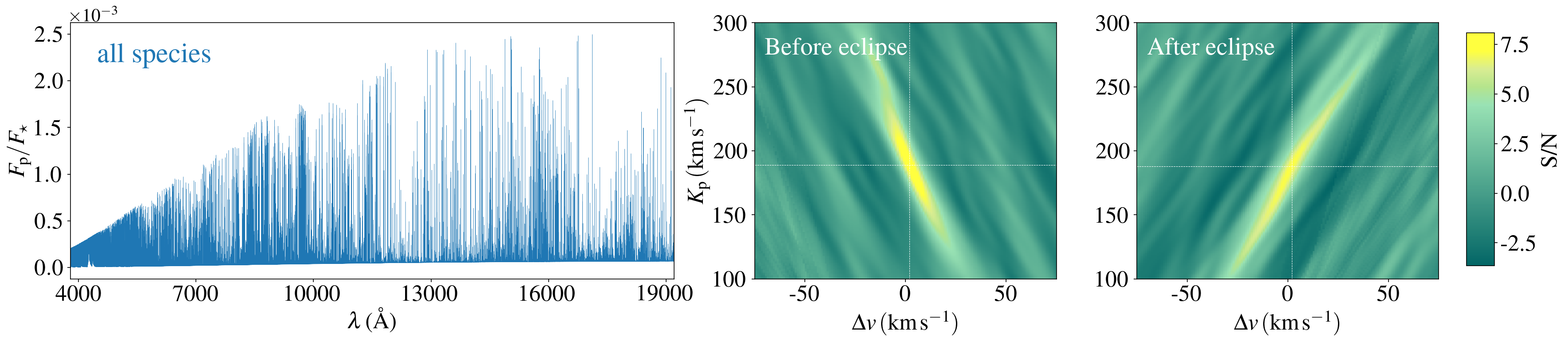}
\caption{\textit{Left panel}: Model spectra obtained by combining all detected species (i.e., \ion{Al}{i}, AlH, \ion{Ca}{ii}, \ion{Cr}{i}, \ion{Fe}{i}, \ion{Fe}{ii}, \ion{Mg}{i}, \ion{Mn}{i}, \ion{Ti}{i}). \textit{Middle and right panels}: S/N maps corresponding to the orbital phase before (night 2) and after (night 1) eclipse, respectively. Dashed white lines mark the locations of the best-fit $K_\mathrm{p}$ and $\Delta v$ values.}
\label{Kpmap_all}
\end{figure*}

To investigate potential atmospheric heterogeneity on \mbox{HAT-P-70b}, we calculated and compared S/N maps with the observation segments obtained before (night 2) and after (night 1) the secondary eclipse, respectively. We computed model spectra of all detected species and subsequently derived S/N maps using the methodology detailed in Sect.~\ref{detection_methods}. The results are presented in Fig.~\ref{Kpmap_all}. Prior to secondary eclipse ingress, the peak of the combined signal was located at $K_\mathrm{p}=188.8_{-24.9}^{+22.8}\,\mathrm{km\,s^{-1}}, \Delta v = 2.2_{-6.4}^{+7.3}\,\mathrm{km\,s^{-1}}$. After eclipse egress, this peak shifted slightly to $K_\mathrm{p}=186.0_{-56.2}^{+45.9}\,\mathrm{km\,s^{-1}}, \Delta v = 1.3\pm19.2\,\mathrm{km\,s^{-1}}$. This lack of significant change indicates an absence of large-scale chemical or thermal heterogeneity across the observed dayside hemisphere. This suggests that the atmospheric dynamics of HAT-P-70b are probably not dominated by a strong equatorial super-rotating jet, but rather by day-to-night flow or weak equatorial winds.

We also reported tentative detections of \ion{C}{i}, \ion{Ca}{i}, \ion{Na}{i}, NaH, and \ion{Ni}{i}. The presence of \ion{C}{i} may result from the thermal dissociation of carbon-bearing molecules such as CO or $\mathrm{CH_4}$ in the high-temperature atmosphere. The \ion{C}{i} signal peaks at $K_\mathrm{p} = 200.7_{-8.4}^{+17.4}\,\mathrm{km\,s^{-1}}$. Although this value is only $13.9\,\mathrm{km\,s^{-1}}$ redshifted from the theoretical expectation, the 1$\sigma$ lower bound is $192.3\,\mathrm{km\,s^{-1}}$, which excludes the theoretical value from the 1$\sigma$ confidence interval. The signals of \ion{Ca}{i} and \ion{Na}{i} were detected only in the CARMENES data but not in the PEPSI observations with higher S/N values. Similarly, the signals of NaH and \ion{Ni}{i} were detected only during the after secondary eclipse phase (night 1), while no significant detection was found during the before secondary eclipse phase (night 2). This discrepancy may be attributed to the higher S/N of the night 1 data (Table~\ref{observe_log}). Therefore, we considered these detections to be tentative.

We also applied the same method to search for AlO, \ion{Ba}{i}, \ion{Ba}{ii}, CaH, \ion{Co}{i}, \ion{Co}{ii}, \ion{Cr}{i}, \ion{Cr}{ii}, CrH, FeH, \ion{K}{i}, \ion{Li}{i}, \ion{Mg}{ii}, MgH, \ion{O}{i}, OH, \ion{Sc}{i}, \ion{Sc}{ii}, \ion{Si}{i}, \ion{Ti}{ii}, TiO, \ion{V}{i}, \ion{V}{ii}, VO, and \ion{Y}{i}. However, no significant signals of these species were detected. Several factors may account for these non-detections. First, some species lack dense or strong spectral lines within the observed wavelength range. For instance, certain metal ions (e.g., \ion{Co}{ii}, \ion{Cr}{ii}, and \ion{V}{ii}) exhibit strong spectral lines just in the ultraviolet band. Second, inaccuracies in the opacity database for some species could hinder effective detection. Third, the S/N of the observational data may be insufficient to reveal weak spectral features. Finally, low atmospheric abundance could also explain the absence of detectable signals. For instance, certain species may undergo thermal dissociation in the hot inversion layer, while others may condense in cooler regions or on the nightside, resulting in their depletion from the observable atmosphere.

%%%%%%%%%%%%%%%%%%%%%%%%%%%%%%%%%%%%%%%%%%%%%%%%%%%%%%%%%%%%%%

\section{Atmospheric retrieval}
\label{atmospheric_retrieval}
\subsection{Retrieval methods}

Based on the detection results with CCF, we performed atmospheric retrievals to constrain the atmospheric properties with the observed thermal emission spectra. We employed the latest retrieval techniques for high-resolution spectroscopy, as described in \citet{yan_2023_CRIRES_wasp}.

The forward models were computed using \texttt{petitRADTRANS}, adopting a free $T$-$P$ profile to describe the atmospheric structure. The $T$-$P$ profile was parameterized by nine free temperature points ($T_8$--$T_0$) uniformly distributed in log-pressure coordinates between $10^{-8}$\,bar (uppermost altitude point, $T_8$) and $10^0$\,bar (lowermost altitude point, $T_0$), with each point set as a free parameter ranging from 100\,K to 6000\,K. Following the methodology of \citet{free_tp}, we imposed a Gaussian prior with a smoothing strength of $\sigma_\mathrm{s}=150\,\mathrm{K\,dex^{-2}}$ to prevent nonphysical zig-zaggy temperature oscillations across pressure scales. The atmospheric structure in \texttt{petitRADTRANS} was divided into 51 discrete layers, uniformly sampled in log-pressure space from the upper boundary at $10^{-8}$\,bar to the lower boundary at $10^0$\,bar. The normalization of models was consistent with the description in Sect.~\ref{detection_methods}.

For the volume mixing ratio (VMR) of the chemical species, we employed two different methods. The first method, chemical free retrieval (FR), assumed that the VMR of each chemical species remained constant across all atmospheric layers. In this case, the logarithm of the VMR for each species, such as $\log(\ion{Al}{i})$ and $\log(\ion{Fe}{i})$, was treated as a free parameter. The second method, chemical equilibrium retrieval (ER), set the relative abundance of each element (e.g., [Al/H], [Fe/H]) as a free parameter, defined by
\begin{equation}
\mathrm{[X/H]} = \log(\mathrm{X_p})-\log(\mathrm{X_\odot}) = \log\left(\mathrm{\frac{X_p}{X_\odot}}\right), 
\end{equation}
where X represents a given element (e.g., Al, Fe), and $\log(\mathrm{X_p})$ and $\log(\mathrm{X_\odot})$  denote the logarithmic VMRs of the element in the planetary and solar compositions, respectively. For the ER method, we utilized the open-source chemical equilibrium code \texttt{pyFastChem} \citep{fastchem_2022} to calculate the VMRs of all species and the mean molecular weight in each atmosphere layer, based on retrieval parameters such as temperature profile and elemental abundance. During chemical equilibrium calculations, we considered 29 elements and the free electrons. The abundances of Al, C, Ca, Cr, Fe, Mg, Mn, Na, Ni, Ti, and V were treated as free parameters, while the abundances of all other elements not targeted in the retrieval were fixed to their solar values. These calculations included 531 chemical species, comprising 501 molecular and ionic species such as CO, FeH, \ion{Ti}{ii}, and \ion{V}{ii}. Condensation processes at low temperatures were excluded from the modelling framework, as they are negligible in the extreme temperature of UHJs while also simultaneously reducing computational cost.

Accurate modelling of spectral line broadening is critical for atmospheric retrieval. The pressure and thermal broadening were already included in the calculation of the opacity linelist. Additionally, we convolved the model spectra with a Gaussian profile corresponding to the resolution of the spectrograph. To account for planetary rotation, we further incorporated rotational broadening using the profile described by \citet{rotational_profile} and convolved it to the model spectra,
\begin{equation}
\begin{split}
& G(x)=\frac{2(1-\varepsilon)(1-x^2)^{\frac{1}{2}} + \frac{\pi\varepsilon}{2}(1-x^2)}{\pi(1-\frac{\varepsilon}{3})},\\
& x=\ln(\frac{\lambda}{\lambda_0})\cdot\frac{c}{v_\mathrm{eq}\sin i_\mathrm{p}},
\label{rotation_eq}
\end{split}
\end{equation}
where $\lambda_0$ is the central wavelength of the spectral line, $\varepsilon$ is the limb darkening coefficient, and $v_\mathrm{eq}$ is the equatorial rotation velocity. We adopted a linear limb darkening model and set the $\varepsilon$ to 1, completely neglecting contributions from the planetary limb to the total radiation flux. Assuming tidal locking for HAT-P-70b, we fixed $v_\mathrm{eq}$ at 3.46\,$\mathrm{km\,s^{-1}}$, consistent with synchronous rotation at the system’s orbital period. To mitigate the smearing effect caused by the planetary orbital motion during the exposure time, the model spectra were further convolved with a box function in velocity space (i.e.,$\ln\lambda$), with the width of the box corresponding to the planetary RV change of approximately 1.5\,$\mathrm{km\,s^{-1}}$ over the course of the \mbox{\textasciitilde5-min} exposure. Furthermore, we applied the fast \texttt{SYSREM} filtering algorithm to the model spectra to obtain the final model spectral matrix, following the methodology outlined in Sect.~\ref{account_sysrem}, with $K_\mathrm{p}$ and $\Delta v$ treated as free parameters. Finally, a Gaussian high-pass filter with a Gaussian $\sigma$ of 31 points was applied to both the final residual and model spectral matrices to remove any remaining broadband features.

We performed atmospheric retrieval by sampling the posterior distribution using the Markov chain Monte Carlo (MCMC) ensemble sampler \texttt{emcee} \citep{emcee}. The logarithm likelihood function was defined as
\begin{equation}
\ln \mathcal{L} = - \sum_{ij} \left[  \frac{(R_{ij} - M_{ij})^2}{(\beta \sigma_{ij})^2} + \ln (2 \pi (\beta \sigma_{ij})^2) \right],
\label{likelihood}
\end{equation}
where $R_{ij}$ and $M_{ij}$ denote the residual and model spectral matrices respectively, $\sigma_{ij}$ represents the residual noise matrix, and $\beta$ are scaling factors accounting for the noise. We ran the MCMC simulation with 20\,000 steps and 200 walkers for each parameter. In the MCMC sampling procedure, the initial 10\,000 steps were discarded as burn-in to ensure adequate convergence of the Markov chains.

\subsection{Retrieval results}

\subsubsection{Basic retrieved parameters}
Our retrievals were performed independently for the FR and ER. The posterior distribution corner maps are presented in Figs.~\ref{corner_FR} and \ref{corner_ER}. 

\begin{figure}
\centering
\includegraphics[width=\hsize]{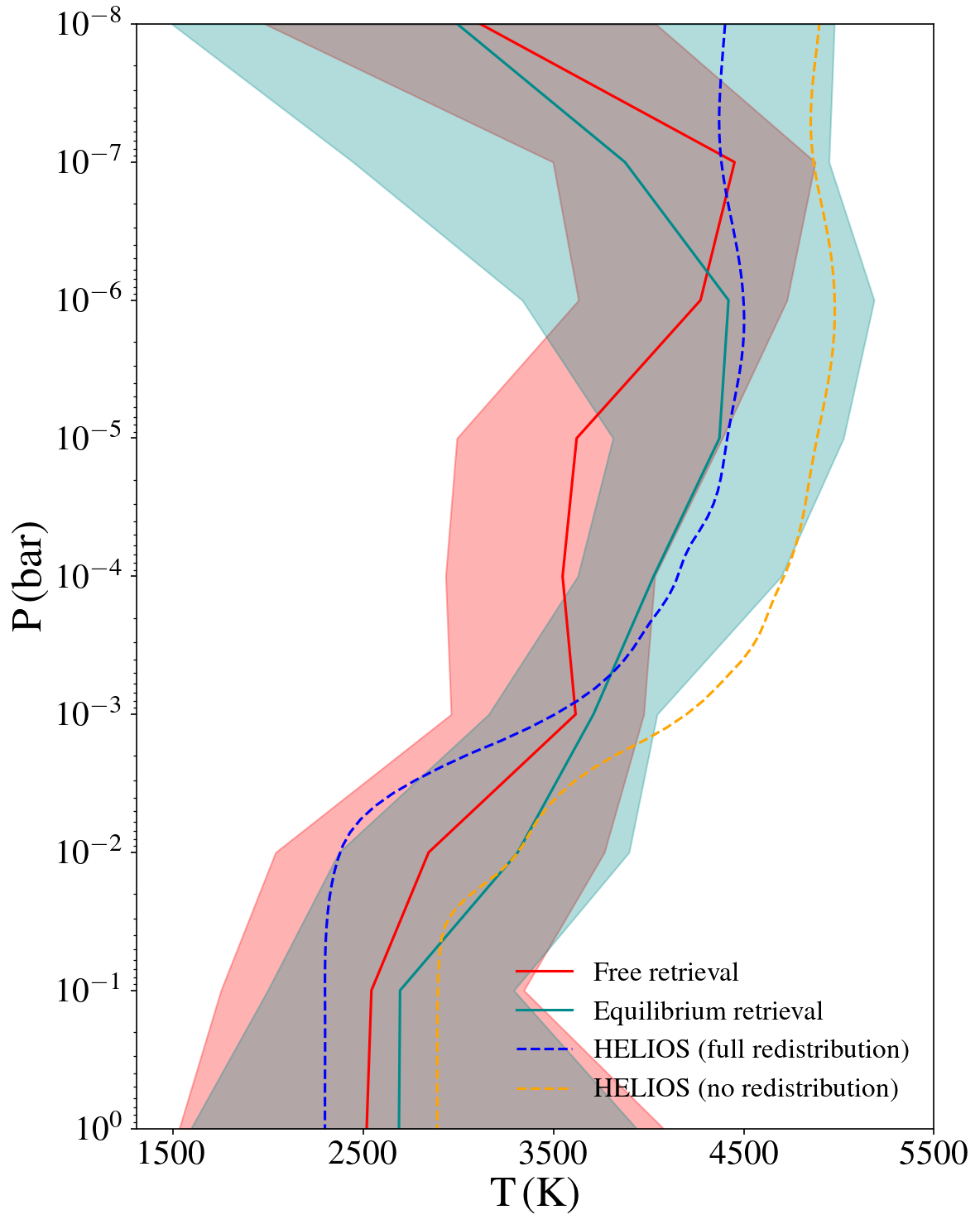}
\caption{Retrieved $T$-$P$ profile (solid line) compared with predictions from the self-consistent \texttt{HELIOS} model (dashed line). The \texttt{HELIOS} model assumes solar elemental abundances and considers two scenarios: no heat redistribution (dayside only) and full heat redistribution between the dayside and nightside. The shaded regions indicate the 1$\sigma$ confidence interval of the retrieved profile.}
\label{retrieval_tp}
\end{figure}

The retrieved $T$-$P$ profiles are shown in Fig.~\ref{retrieval_tp}. The ER results indicate the presence of a temperature inversion layer between $10^{-1}$ and $10^{-5}$\,bar, with a temperature difference of approximately 1700\,K. In contrast, the FR results suggest a more complex, double-inversion structure. The lower inversion layer ($10^{-1}$ to $10^{-3}$\,bar) is likely retrieved from the neutral metal lines such as \ion{Fe}{i}, while the higher inversion layer ($10^{-5}$ to $10^{-7}$\,bar) is plausibly retrieved from ionized metal lines (e.g., \ion{Fe}{ii} and \ion{Ca}{ii}). Additionally, we employed a modified version of the \texttt{HELIOS} code \citep{helios} to compute a self-consistent atmospheric model. A detailed description of the \texttt{HELIOS} model calculation can be found in \citet{kelt-20_yan}. Due to the relatively large uncertainties in the retrieved profile, the \texttt{HELIOS} model shows agreement with the retrieval results, except in the uppermost atmosphere ($10^{-7}$ to $10^{-8}$\,bar), where discrepancies are more pronounced.

Both $K_\mathrm{p}$ and $\Delta v$ were included as free parameters in all retrievals. The FR yields a $K_\mathrm{p}$ value of $187.9^{+1.2}_{-1.1}\,\mathrm{km\,s^{-1}}$, while the ER result is $187.7^{+1.2}_{-1.1}\,\mathrm{km\,s^{-1}}$, both consistent with the values obtained from CCF. Although the best-fit values show a slight deviation of \textasciitilde\,$1.0\,\mathrm{km\,s^{-1}}$ relative to the theoretical expectation, these offsets lie within the 1$\sigma$ uncertainty range. The retrieved $\Delta v$ is $2.19^{+0.37}_{-0.40},\mathrm{km\,s^{-1}}$ for FR and $2.15^{+0.37}_{-0.36},\mathrm{km\,s^{-1}}$ for ER, both indicating a redshift of approximately $2.0\,\mathrm{km\,s^{-1}}$. This redshift may be attributed to day-to-night atmospheric winds on the planet, although uncertainties of $v_\mathrm{sys}$ and transit ephemeris can also contribute to the $\Delta v$ shift \citep{offset_parameters}.

For the metal abundance, the FR yielded a \ion{Fe}{i} VMR of $\mathrm{log(\ion{Fe}{i})} = -4.16^{+0.74}_{-1.11}$, corresponding to a metallicity of $\mathrm{[Fe/H]} = 0.38^{+0.74}_{-1.11}$. In contrast, the ER gave a metallicity of $\mathrm{[Fe/H]} = 0.23^{+1.08}_{-0.98}$. Both retrieved metallicity values are consistent with the solar value within the $1\sigma$ range. 

\subsubsection{Elemental abundance pattern}
\begin{figure*}
\centering
\includegraphics[width=\textwidth]{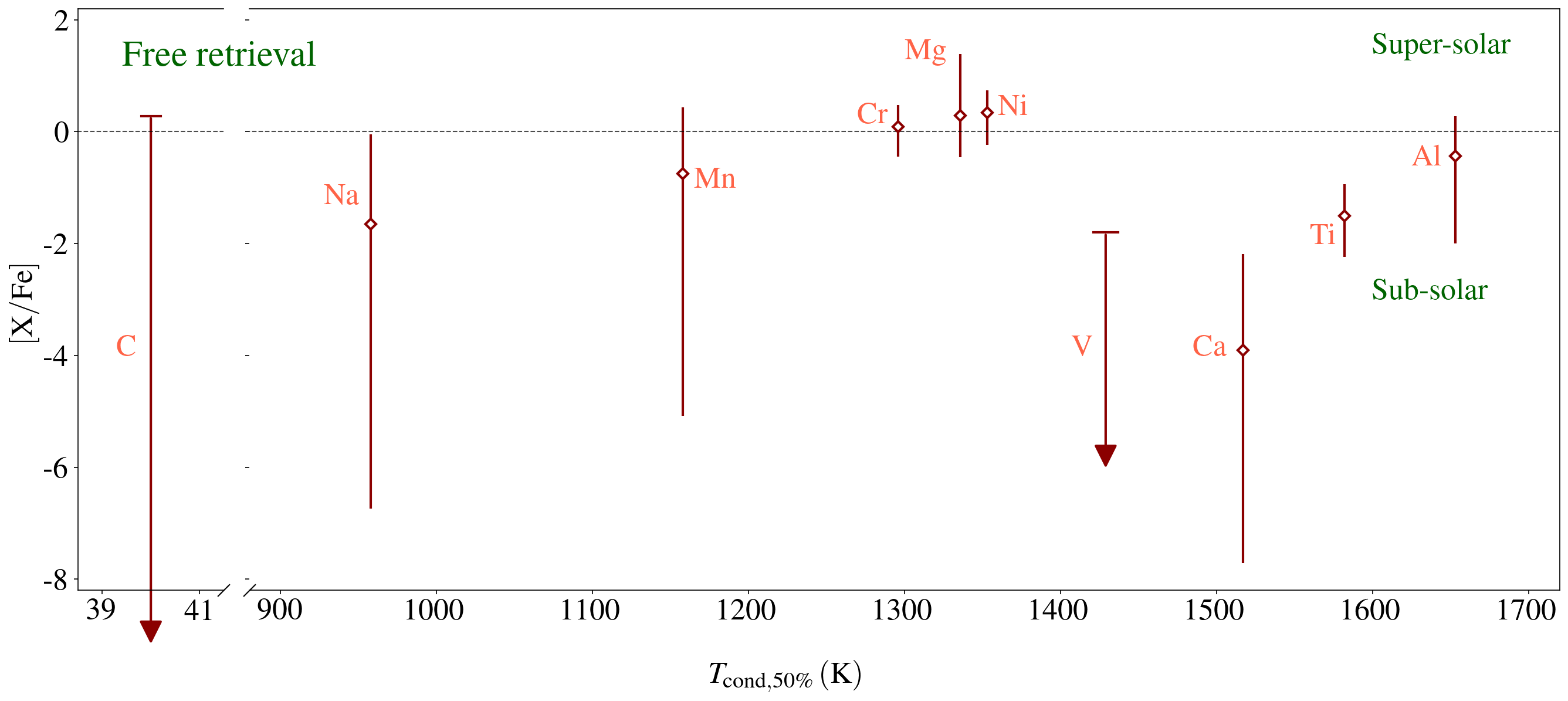}
\caption{Elemental abundances in the atmosphere of HAT-P-70b relative to iron, derived from chemical free retrieval, as a function of condensation temperature. The dashed black line represents consistency with solar abundance. All error bars indicate 1$\sigma$ uncertainties, while the abundances of C and V are reported as upper limits.}
\label{abun_FR}
\end{figure*}

\begin{figure*}
\centering
\includegraphics[width=\textwidth]{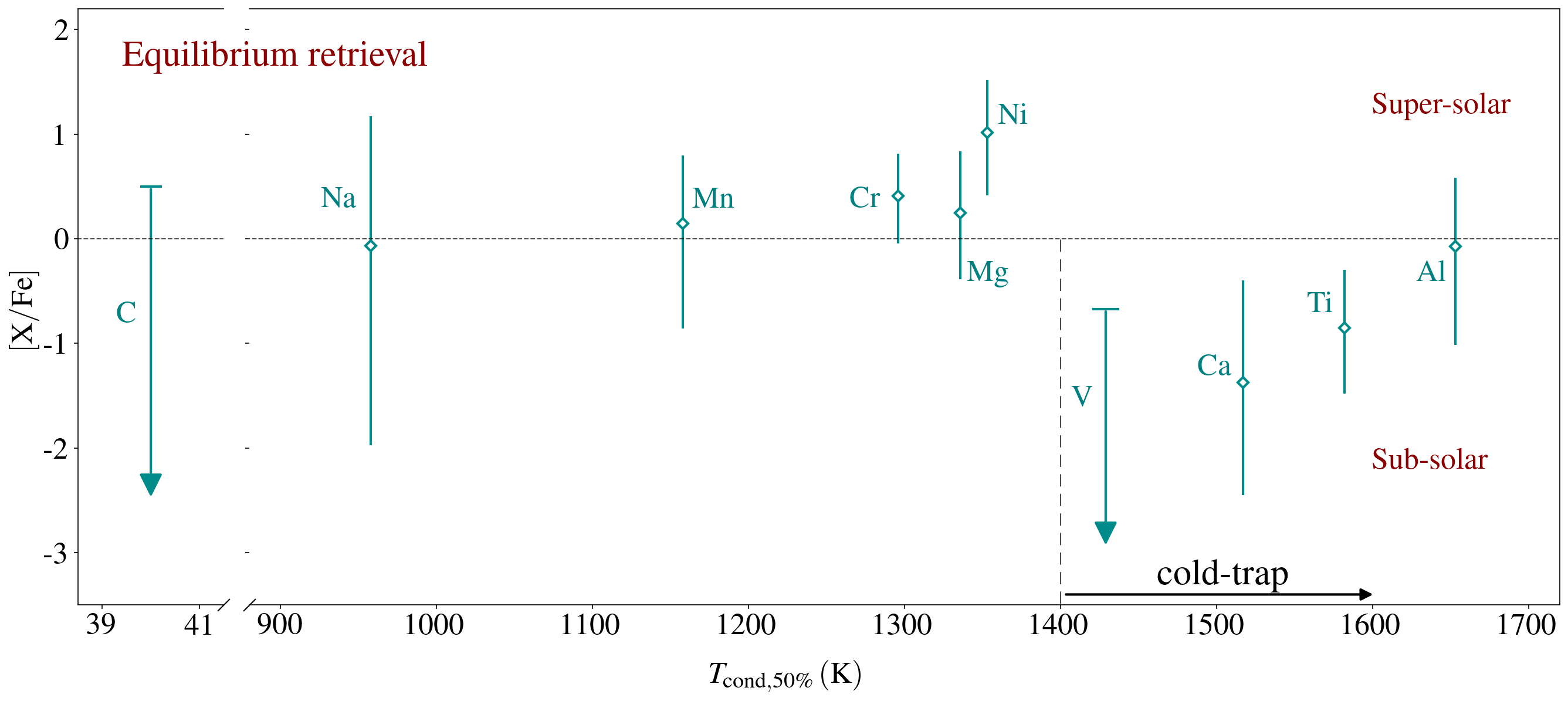}
\caption{Same as Fig.~\ref{abun_FR} but for chemical equilibrium retrieval results.}
\label{abun_CE}
\end{figure*}

From the MCMC posterior samples, we derived the solar-relative abundance for each element ([X/H]) and subsequently converted them into iron-relative abundances ([X/Fe]):
\begin{equation}
[\mathrm{X/Fe}] = \log\left(\mathrm{\frac{X_p}{X_\odot}}\right) - \log\left(\mathrm{\frac{Fe_p}{Fe_\odot}}\right) = \mathrm{[X/H]} - \mathrm{[Fe/H]}.
\end{equation}
These results are presented in Fig.~\ref{abun_FR} for FR and Fig.~\ref{abun_CE} for ER, with the specific values summarized in Table~\ref{metal_abun_table}. In the FR results, the abundance of Al was determined by both \ion{Al}{i}
and AlH, whereas the abundances of all other elements were inferred solely
from their atomic species, implying an underlying assumption that these elements
are predominantly in atomic form. 

The retrieved abundance of Na appears slightly subsolar, likely due to thermal dissociation and ionization at high temperatures. This element are highly prone to ionization and their ionic forms may dominate in the atmosphere, causing the atomic-based retrievals to significantly underestimate their true abundances. V was not detected, with the retrieval providing only a $1\sigma$ upper limit of $\mathrm{log(V)}<-9.69$, which is substantially below the solar value ($\log (\mathrm{V})_\odot=-8.04$). For Ti, its high condensation temperature suggests potential condensation on the planet’s cooler nightside. However, efficient atmospheric circulation likely transports Ti-bearing species to the dayside hemisphere of HAT-P-70b, resulting in only a mild depletion with a Ti abundance  slightly below solar. The abundances of all other elements are consistent with solar values within the $1\sigma$ range. 

Based on the high-resolution transmission spectra observed by HARPS-N, \citet{hatp70b_abun_harpsn} also retrieved the abundances of various metal species in the atmosphere of HAT-P-70b using the FR method. Most of the retrieved abundances show a good agreement within 1$\sigma$ between the two works. Only the abundance of Cr shows notable difference, but the values are still within the 2$\sigma$ uncertainty range.

Our ER analysis, which accounts for thermal ionization, yields a Na abundance consistent with the solar value. Similarly, the abundances of Mn, Cr, and Mg also align with solar abundances, corroborating the findings from our FR results. Among all retrieved elements, Ni is the only one showing a significantly super-solar abundance, with $\mathrm{[Ni/Fe]} = 1.01^{+0.50}_{-0.60}$. This Ni enrichment, also observed on another UHJ, namely \mbox{WASP-76b} \citep{wasp76b_cold_trap}, may be attributed to the accretion of Ni-rich planetesimals during the planet's formation. For refractory elements with high condensation temperatures ($T_\mathrm{cond,50\%} > 1400\,\mathrm{K}$), such as V, Ca, and Ti, the inferred abundances are modestly depleted, consistent with the FR findings. In the case of Al, which has the highest condensation temperature among the retrieved elements, no significant depletion is observed. This may be due to disequilibrium processes, with a substantial fraction of Al stored in molecular form (e.g., AlH). Alternatively, the planet may have formed with an inherent enrichment of Al (similarly to Ni) and despite being potentially depleted due to cold-trapping, the dayside hemispheric Al abundance could still appear consistent with the solar value as a result.

\begin{figure}
\centering
\includegraphics[width=\hsize]{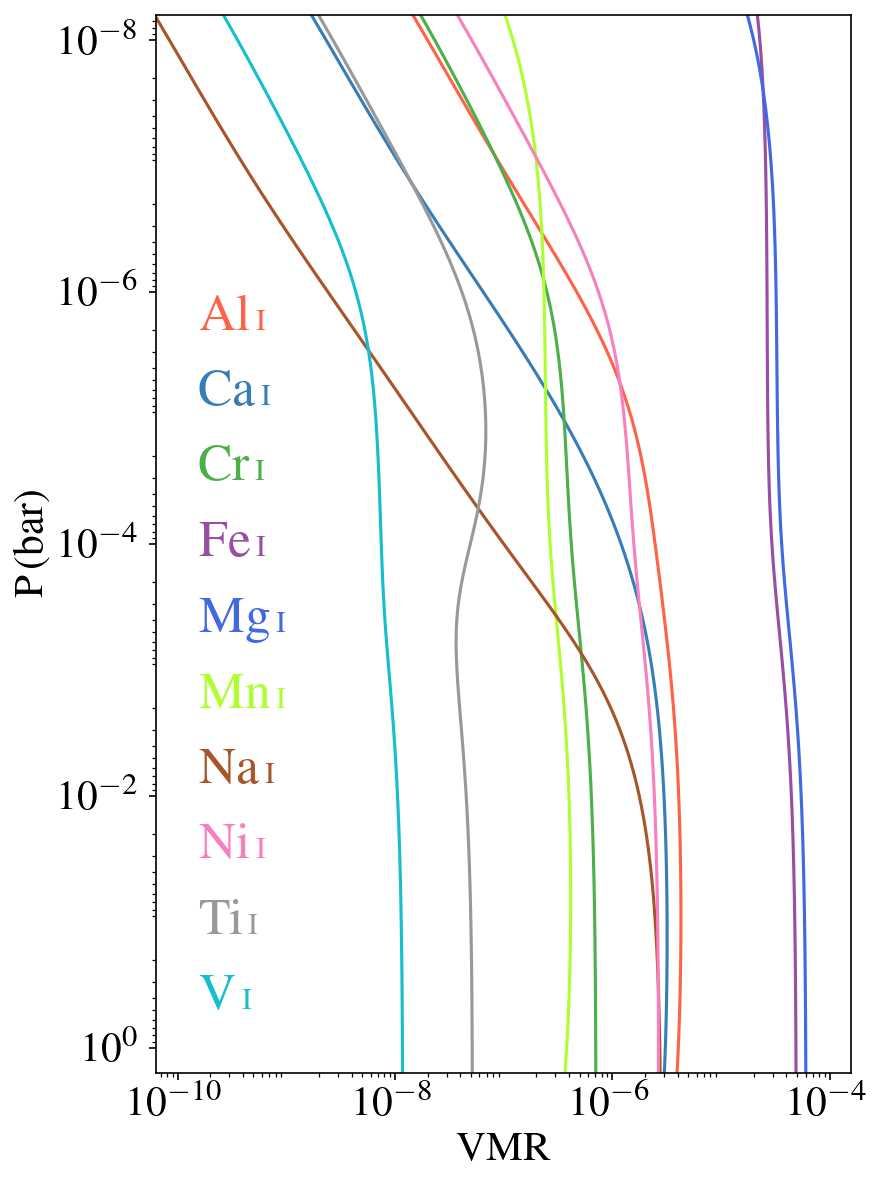}
\caption{Volume mixing ratios of all targeted metal atoms as a function of pressure at a fixed isothermal temperature of 2500\,K under the assumption of chemical equilibrium. All elemental abundances are assumed to be solar. The results are illustrative calculations intended to demonstrate metal ionization, rather than retrieval results.}
\label{VMR_metal}
\end{figure}

Compared to the FR results, the uncertainties on the abundances derived from the ER are smaller, suggesting that the chemical equilibrium assumption more accurately represents the atmospheric condition of HAT-P-70b. Furthermore, the ER consistently yields higher abundances for most elements compared to the FR (Table~\ref{metal_abun_table}), which is a natural result of metal ionization. As shown in Fig.~\ref{VMR_metal}, the VMRs of most metals significantly drop towards lower pressures due to strong ionization in the upper planetary atmosphere. Therefore, the free retrieval, which only accounts for atomic species, tends to underestimate the abundances. The only exception is Mg, which has similar retrieved abundances between ER and FR. This is not surprising because Mg and Fe are known to be less prone to ionization, as suggested by chemical equilibrium calculations (Fig.~\ref{VMR_metal}).

We also attempted to constrain the abundance of C using the tentative detection of \ion{C}{i}. The FR results suggested an upper limit of log(\ion{C}{i}) < -3.09 (corresponding to [C/H] < 0.56). The ER yielded a tighter constraint of [C/H] < 0.74. Since the retrieval cannot constrain well the abundance and only an upper limit is yielded, we considered the tentative \ion{C}{i} signal as a possible false positive.

It is also important to note that the detection of a species in the planetary atmosphere does not necessarily imply that the corresponding element is not depleted. For example, although we detected a $\mathrm{S/N}\sim5.3$ emission signal for \ion{Ti}{i} with CCF, the retrievals still yielded a subsolar Ti abundance (FR: $\mathrm{[Ti/Fe]} = -1.51^{+0.56}_{-0.74}$, ER: $\mathrm{[Ti/Fe]} = -0.85^{+0.55}_{-0.63}$). This highlights that a definitive assessment of elemental depletion should rely on quantitative abundance retrieval rather than detection significance alone.

\subsubsection{Evidence of disequilibrium processes}

\begin{figure}
\centering
\includegraphics[width=\hsize]{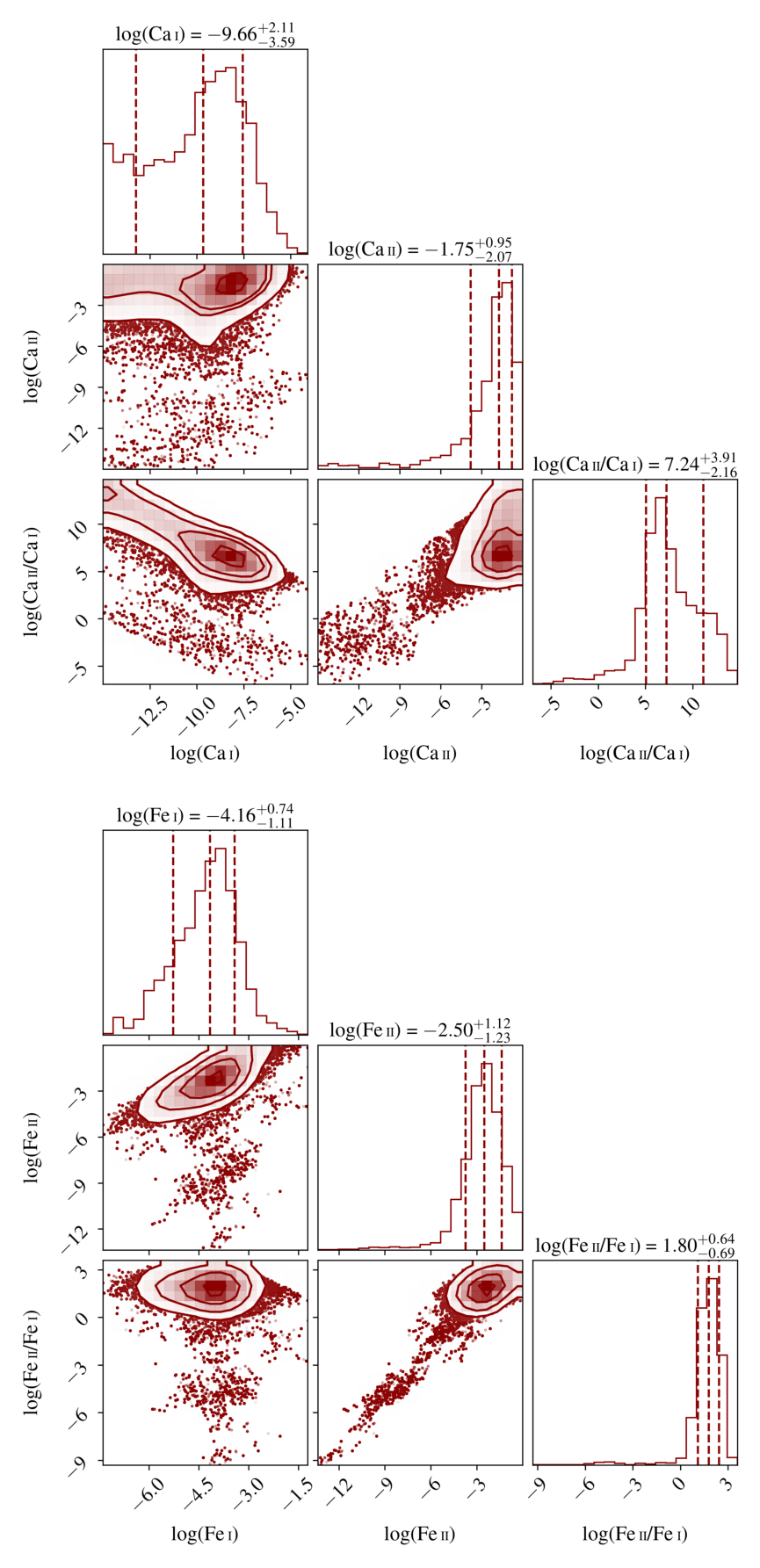}
\caption{Posterior distribution corner plots for the VMRs of \ion{Ca}{i} and \ion{Ca}{ii} (top), and \ion{Fe}{i} and \ion{Fe}{ii} (bottom), obtained from the chemical free retrieval results. The relative abundances between these species are also calculated.}
\label{vmr_corner}
\end{figure}

Our FR independently constrained the abundances of each species, revealing some results inconsistent with chemical equilibrium models. Notably, the retrieved VMRs for ionized species were unexpectedly high: $\mathrm{log(\ion{Ca}{ii})}=-1.75^{+0.95}_{-2.07}$ and $\mathrm{log(\ion{Fe}{ii})}=-2.50^{+1.12}_{-1.23}$. The relative abundances between atomic and ionic species were also calculated. The results showed that the VMR of \ion{Ca}{ii} exceeds that of \ion{Ca}{i} by approximately seven orders of magnitude ($\mathrm{log(\ion{Ca}{ii}/\ion{Ca}{i})}=7.24^{+3.91}_{-2.16}$), while the VMR of \ion{Fe}{ii} is about two orders of magnitude greater than that of \ion{Fe}{i} ($\mathrm{log(\ion{Fe}{ii}/\ion{Fe}{i})}=1.80^{+0.64}_{-0.69}$). The posterior distributions for these species are shown in Fig.~\ref{vmr_corner}.

These discrepancies may arise from some disequilibrium processes, predominately the hydrodynamic expanding effect. In the above retrievals, we assumed the atmosphere to be hydrostatic. However, UHJs could have strong atmospheric escape with outward-expanding flow \citep{wasp-33_yan_2019, Sing2019}. This type of hydrodynamic process significantly enhances the atmospheric density at high altitudes. Since the detected lines of the ionized species are likely formed at high altitudes, this hydrodynamic effect is more prominent for ionic lines compared to the atomic lines. Thereby, our retrieval under hydrostatic assumption tends to compensate this effect by enhancing the VMRs of \ion{Ca}{ii} and \ion{Fe}{ii} to extremely exaggerated values. Further atmosphere modeling work that would include hydrodynamic escape could help resolve the discrepancies and provide a credible mass loss rate \citep{Huang2023}.

Furthermore, our analysis relies on forward models calculated under the assumption of local thermodynamic equilibrium (LTE), which neglects potential non-LTE (NLTE) effects. As demonstrated by \citet{inversion_theory4_kelt9b_nlte}, NLTE effects can alter the population distribution of atomic energy levels for metals (e.g., Fe and         Mg) in the atmospheres of UHJs, thereby impacting both the atmospheric thermal structure and the strength of spectral lines. For example, NLTE effects may enhance certain spectral features, such as \ion{Fe}{ii} lines, beyond the predictions of an LTE model. Consequently, an atmospheric retrieval framework built upon LTE assumptions may erroneously compensate for these stronger-than-expected lines by overestimating the VMRs of metal species to fit the observed spectra.

\section{Conclusions}
\label{conclusions}

We observed the dayside thermal emission spectra of the UHJ HAT-P-70b with the high-resolution spectrographs CARMENES and PEPSI, both covering orbital phases before and after the secondary eclipse. Atmospheric features were identified using the CCF method, followed by atmospheric retrievals to constrain the elemental abundances and the dayside $T$-$P$ profile. Our main conclusions are summarized below:

\begin{itemize}
\item[--] We detected the emission signals of \ion{Al}{i}, AlH, \ion{Ca}{ii}, \ion{Cr}{i}, \ion{Fe}{i}, \ion{Fe}{ii}, \ion{Mg}{i}, \ion{Mn}{i}, and \ion{Ti}{i}. This marks the first detection of \ion{Al}{i} and AlH in an exoplanet's atmosphere. We also presented tentative detections of \ion{C}{i}, \ion{Ca}{i}, \ion{Na}{i}, NaH, and \ion{Ni}{i}.

\item[--] Given the high condensation temperature of aluminum and titanium, the detections of \ion{Al}{i}, AlH, and \ion{Ti}{i} imply efficient energy and materials transport between the day and night side hemispheres of HAT-P-70b. Furthermore, our findings suggest that AlH is a more prominent aluminum reservoir than AlO in UHJ atmospheres, highlighting its importance for future atmospheric characterization efforts.

\item[--] A combined CCF analysis incorporating all detected species revealed no significant velocity offset between before and after the secondary eclipse observations, suggesting no strong longitudinal heterogeneity. This suggests that the atmospheric dynamics of HAT-P-70b are likely not dominated by a strong equatorial super-rotating jet.

\item[--] We independently performed FR and ER to constrain the planet’s atmospheric properties. Both retrieval frameworks consistently identified a strong thermal inversion in the dayside hemisphere, with a temperature increase of \textasciitilde\,1500–2000\,K between the lower and the upper planetary atmosphere.

\item[--] The retrieved metallicity [Fe/H] is $0.38^{+0.74}_{-1.11}$ from FR and $0.23^{+1.08}_{-0.98}$ from ER. The abundances of Na, Mn, Cr, and Mg are consistent with solar values, whereas those of V, Ca, and Ti appear moderately depleted, likely as a consequence of cold-trapping on the planet’s cooler nightside, owing to their high condensation temperatures ($T_\mathrm{cond,50\%} > 1400\,\mathrm{K}$). Notably, Ni shows a tentatively mild enrichment, which may be attributed to the accretion of Ni-rich planetesimals during the planet’s formation and evolution. In particular, Al, despite having the highest condensation temperature, exhibits an abundance consistent with the solar value, which could be due to its initial enrichment during formation.

\item[--] The FR analysis, which independently constrained the VMRs of atoms and ions, revealed significantly high abundances of ionized species such as \ion{Fe}{ii} and \ion{Ca}{ii}. These results strongly deviate from chemical equilibrium predictions under hydrostatic assumption, indicating the high altitude atmosphere is likely undergoing significant hydrodynamic escaping.

\end{itemize}

In this work, we present the first dayside observation of HAT-P-70b, with the discovery of a rich array of spectral features characterizing refractory species. The elemental abundance pattern of these detected species offers important insight into the planet’s formation and evolutionary history. However, our observations were primarily conducted in the visible band and lack a high-quality IR data to detect volatile species such as CO. As a result, the C/O ratio remains poorly constrained. Future observations using ground-based high-resolution IR spectrographs (e.g., CRIRES+, IGRINS, IGRINS-2) or space-based facilities such as JWST could provide a more comprehensive characterization of the planet’s elemental abundance pattern.

%%%%%%%%%%%%%%%%%%%%%%%%%%%%%%%%%%%%%%%%%%%%%%%%%%%%%%%%%%%%%%
\begin{acknowledgements}
We acknowledge the support by the National Natural Science Foundation of China (grant no. 42375118). 
This publication was based on observations collected under the CARMENES Legacy+ project. 
CARMENES is an instrument at the Centro Astron\'omico Hispano en Andaluc\'ia (CAHA) at Calar Alto (Almer\'{\i}a, Spain), operated jointly by the Junta de Andaluc\'ia and the Instituto de Astrof\'isica de Andaluc\'ia (CSIC).
Funding for CARMENES has been provided by the
Max-Planck-Institut f\"ur Astronomie (MPIA),
Consejo Superior de Investigaciones Cient\'{\i}ficas (CSIC),
European Regional Development Fund (ERDF),
Ministerio de Ciencia, Innovaci\'on y Universidades (MICIU),
Deutsche Forschungsgemeinschaft (DFG)
and the members of the CARMENES Consortium
(\url{https://carmenes.caha.es}).
The LBT is an international collaboration among institutions in the United States, Italy and Germany. LBT Corporation partners are: The University of Arizona on behalf of the Arizona Board of Regents; Istituto Nazionale di Astrofisica, Italy; LBT Beteiligungsgesellschaft, Germany, representing the Max-Planck Society, The Leibniz Institute for Astrophysics Potsdam, and Heidelberg University; The Ohio State University, representing OSU, University of Notre Dame, University of Minnesota and University of Virginia. We thanks the German Federal Ministry (BMBF) for the year-long support for the construction of PEPSI through their Verbundforschung grants 05AL2BA1/3 and 05A08BAC as well as the State of Brandenburg for the continuing support of PEPSI for the LBT (\url{https://pepsi.aip.de/}). 
We acknowledge financial support from the Agencia Estatal de Investigaci\'on (AEI/10.13039/501100011033) of the MICIU and the ERDF ``A way of making Europe'' through projects
 PID2022-137241NB-C4[1:4],      % IAC1+CAB+IAA+UCM
 PID2021-125627OB-C3[1:2],      % ICE+IAC2
 PID2021-126365NB-C21,
and the Centre of Excellence ``Severo Ochoa'' and ``Mar\'ia de Maeztu'' awards to the Instituto de Astrof\'isica de Andaluc\'ia (CEX2021-001131-S) and Institut de Ci\`encies de l'Espai (CEX2020-001058-M).
D.C. is supported by the LMU-Munich Fraunhofer-Schwarzschild Fellowship and by the Deutsche Forschungsgemeinschaft (DFG, German Research Foundation) under Germany's Excellence Strategy -- EXC 2094 -- 390783311.
This work was co-funded by the European Union (ERC-CoG, EVAPORATOR, Grant agreement No. 101170037). Views and opinions expressed are however those of the author(s) only and do not necessarily reflect those of the European Union or the European Research Council. Neither the European Union nor the granting authority can be held responsible for them.
\end{acknowledgements}

% Please note that we have included the references below in
% order to compile the document, but we ask you to:
\bibliographystyle{aa} % style aa.bst
\bibliography{ref.bib} % your references Yourfile.bib

% %%%%%%%%%%%%%%%%%%%%%%%%%%%%%%%%%%%%%%%%%%%%%%%%%%%%%%%%%%%%%%
% Example below of non-structurated natbib references  
% To use the v8.3 macros with this form of composition of bibliography,
% the option "bibyear" should be added to the command line
% "\documentclass[bibyear]{aa}".
% %%%%%%%%%%%%%%%%%%%%%%%%%%%%%%%%%%%%%%%%%%%%%%%%%%%%%%%%%%%%%%

%%%%%%%%%%%%%%%%%%%%%%%%%%%%%%%%%%%%%%%%%%%%%%%%%%%%%%%%%%%%%%%
% Appendices must be placed after   \end{thebibliography}
% They will be placed automatically on a new page.
%%%%%%%%%%%%%%%%%%%%%%%%%%%%%%%%%%%%%%%%%%%%%%%%%%%%%%%%%%%%%%%
\begin{appendix}
\section{Additional tables and figures}

\begin{figure*}[htbp]
\centering
\includegraphics[width=\hsize]{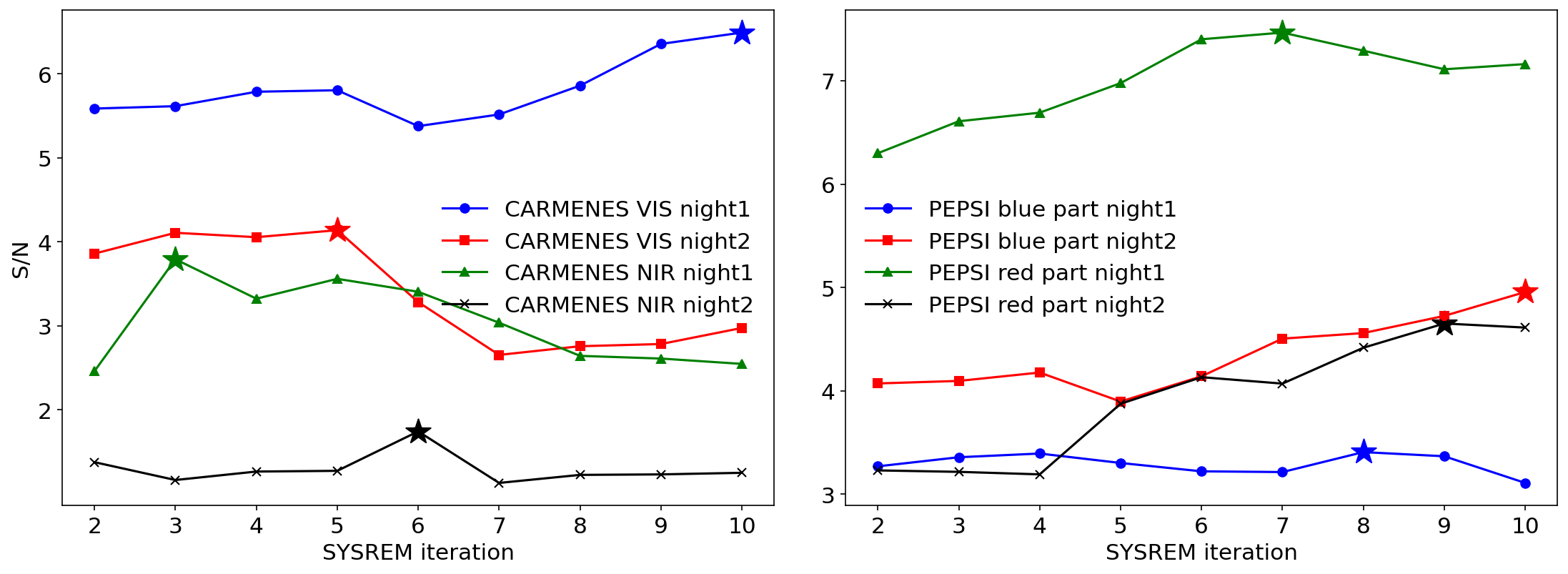}
\caption{S/N values of the \ion{Fe}{i} signal as a function of the \texttt{SYSREM} iteration. The highest S/N values are marked with stars. Each S/N value is measured at the peak position corresponding to the strongest detection signal.}
\label{SNR_iter}
\end{figure*}

\begin{table*}
\centering
\caption{Opacity databases employed for CCF calculation and atmospheric retrieval.}
\label{opacity}
\begin{tabular}{lll}
\hline \hline
Chemical species & Opacity & References \\
\hline
\ion{Al}{i}, \ion{Ba}{i}, \ion{Ba}{ii}, \ion{C}{i}, \ion{Ca}{i}, \ion{Ca}{ii}, \ion{Cr}{i}, \ion{Cr}{ii}, \ion{Fe}{i}, \ion{Fe}{ii}, & Kurucz & \citet{kurucz_opacity} \\
\ion{Li}{i}, \ion{Mg}{i}, \ion{Mg}{ii}, \ion{Na}{i}, \ion{Ni}{i}, \ion{Si}{i}, \ion{Ti}{i}, \ion{Ti}{ii}, \ion{Y}{i} \\
\ion{Co}{i}, \ion{Co}{ii}, \ion{Mn}{i}, \ion{Sc}{i}, \ion{Sc}{ii}, \ion{V}{i}, \ion{V}{ii} & VALD & \citet{vald_opacity1, vald_opacity2}\\
CaH, CrH, FeH, MgH, OH & MOLLIST & \citet{mollist_opacity, cah_opacity, mgh_opacity}; \\
 & & \citet{crh_opacity, oh_opacity1, oh_opacity2}\\
\ion{K}{i} & Allard & \citet{pRT_2019}\\
AlH & AlHambra & \citet{alh_opacity}\\
AlO & ExoMol & \citet{alo_opacity}\\
NaH & Rivlin & \citet{nah_opacity}\\
TiO & Toto & \citet{tio_opacity}\\
VO & VOMYT & \citet{vo_opacity}\\
\hline
\end{tabular}
\tablefoot{Opacity data for AlH, \ion{Ba}{i}, \ion{Ba}{ii}, \ion{Co}{i}, \ion{Co}{ii}, \ion{Cr}{i}, CrH, \ion{Mn}{i}, NaH, \ion{Ni}{i}, \ion{Sc}{i}, \ion{Sc}{ii}, \ion{V}{i}, and \ion{V}{ii} were downloaded from the \href{https://dace.unige.ch}{DACE}, while the remaining species were obtained via \href{https://keeper.mpdl.mpg.de/d/ccf25082fda448c8a0d0}{petitRADTRANS opacity database}.}
\end{table*}

\begin{table*}
\centering
\caption{Summary of the retrieved elemental abundances.}
\label{metal_abun_table}
{\renewcommand{\arraystretch}{1.2}
\begin{tabular}{llccccc}
\hline
\hline
% \noalign{\smallskip}
Element & FR: log(VMR) & FR: [X/H] & FR: [X/Fe] & ER: [X/H] & ER: [X/Fe] & $T_\mathrm{cond,50\%}\,\mathrm{(K)}$ \\
% \noalign{\smallskip}
\hline
% \noalign{\smallskip}
C & $ < -3.09$ & $ < 0.56$ & $ < 0.28$ & $ < 0.74$ & $ < 0.50$ & 40 \\
% \noalign{\smallskip}
Na & $-7.42_{-4.84}^{+1.93}$ & $-1.64_{-4.84}^{+1.93}$ & $-1.65_{-5.09}^{+1.60}$ & $0.03_{-1.81}^{+1.42}$ & $-0.07_{-1.90}^{+1.24}$ & 958 \\
% \noalign{\smallskip}
Mn & $-7.36_{-4.02}^{+1.65}$ & $-0.78_{-4.02}^{+1.65}$ & $-0.76_{-4.33}^{+1.19}$ & $0.19_{-1.21}^{+1.26}$ & $0.15_{-1.01}^{+0.65}$ & 1158 \\
% \noalign{\smallskip}
Cr & $-5.97_{-1.32}^{+0.92}$ & $0.41_{-1.32}^{+0.92}$ & $0.09_{-0.53}^{+0.39}$ & $0.56_{-1.02}^{+1.17}$ & $0.41_{-0.46}^{+0.40}$ & 1296 \\
% \noalign{\smallskip}
Fe & $-4.16_{-1.11}^{+0.74}$ & $0.38_{-1.11}^{+0.74}$ & 0 & $0.23_{-0.98}^{+1.08}$ & 0 & 1334\\
% \noalign{\smallskip}
~ & $-2.50^{+1.12}_{-1.23}$ (\ion{Fe}{ii}) \\
% \noalign{\smallskip}
Mg & $-3.88_{-1.05}^{+1.34}$ & $0.57_{-1.05}^{+1.34}$ & $0.28_{-0.74}^{+1.11}$ & $0.46_{-1.22}^{+1.19}$ & $0.25_{-0.63}^{+0.59}$ & 1336 \\
% \noalign{\smallskip}
Ni & $-5.20_{-1.22}^{+0.91}$ & $0.60_{-1.22}^{+0.91}$ & $0.34_{-0.58}^{+0.40}$ & $1.18_{-1.02}^{+1.13}$ & $1.01_{-0.60}^{+0.50}$ & 1353 \\
% \noalign{\smallskip}
V & $ < -9.69$ & $ < -1.59$ & $ < -1.80$ & $ < 0.52$ & $ < -0.67$ & 1429 \\
% \noalign{\smallskip}
Ca & $-9.66_{-3.59}^{+2.11}$ & $-3.96_{-3.59}^{+2.11}$ & $-3.91_{-3.80}^{+1.73}$ & $-1.39_{-1.07}^{+1.48}$ & $-1.37_{-1.07}^{+0.98}$ & 1517 \\
% \noalign{\smallskip}
~ & $-1.75^{+0.95}_{-2.07}$ (\ion{Ca}{ii}) \\
% \noalign{\smallskip}
Ti & $-8.36_{-1.17}^{+0.96}$ & $-1.33_{-1.17}^{+0.96}$ & $-1.51_{-0.74}^{+0.56}$ & $-0.73_{-0.91}^{+1.08}$ & $-0.85_{-0.63}^{+0.55}$ & 1582 \\
% \noalign{\smallskip}
Al & $-5.85_{-1.88}^{+1.15}$ & $-0.28_{-1.88}^{+1.15}$ & $-0.44_{-1.56}^{+0.71}$ & $-0.04_{-1.07}^{+1.19}$ & $-0.08_{-0.94}^{+0.66}$ & 1653 \\
% \noalign{\smallskip}
~ & $-7.00^{+1.51}_{-4.37}$ (\ion{Al}{i}) \\
% \noalign{\smallskip}
~ & $-7.09^{+2.24}_{-5.08}$ (AlH) \\
% \noalign{\smallskip}
\hline
\end{tabular}}
\tablefoot{FR and ER denote the chemical free and chemical equilibrium retrieval, respectively. Except for aluminum, whose abundance was determined using both \ion{Al}{i} and AlH, the abundances of all other elements were derived exclusively from their atomic species.}
\end{table*}

\begin{figure*}
\centering
\includegraphics[width=\textwidth]{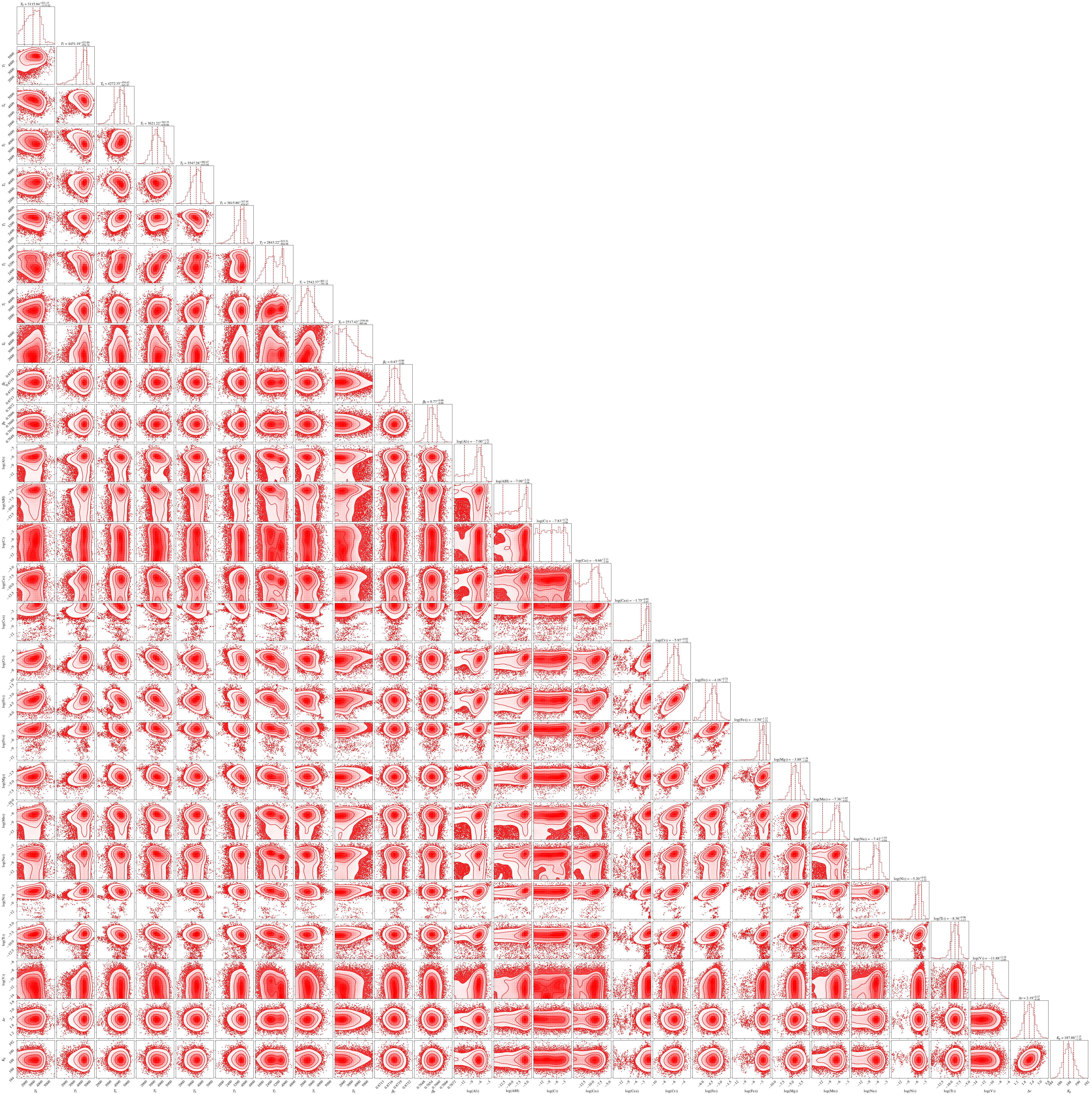}
\caption{Posterior distribution of the parameters from the chemical free retrieval.}
\label{corner_FR}
\end{figure*}

\begin{figure*}
\centering
\includegraphics[width=\textwidth]{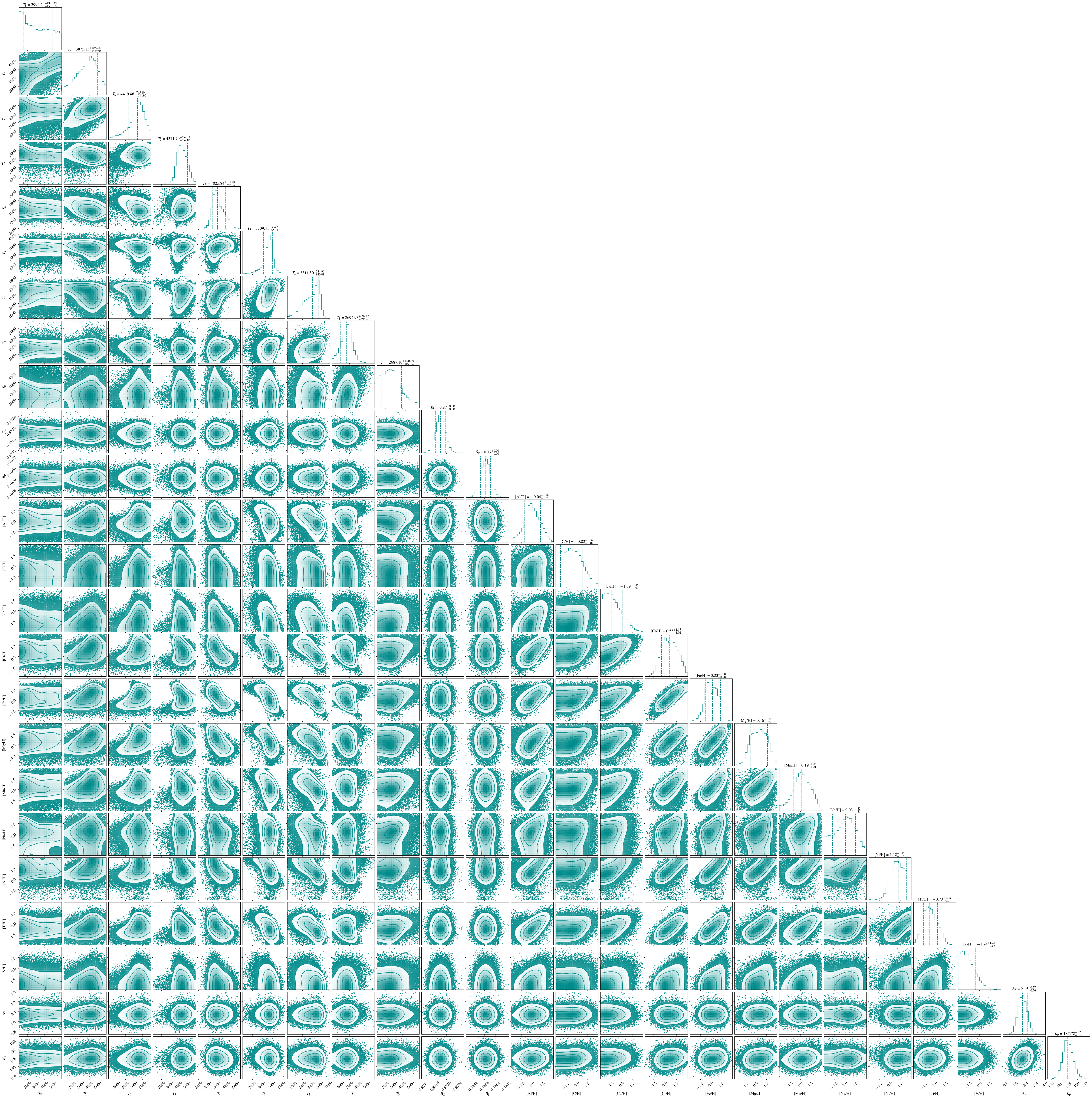}
\caption{Posterior distribution of the parameters from the chemical equilibrium retrieval.}
\label{corner_ER}
\end{figure*}

\end{appendix}

\end{document}